\documentclass[letterpaper,twocolumn,10pt]{article}
\pdfoutput=1
\usepackage{usenix,epsfig,endnotes,subcaption,graphicx}
\usepackage{color}
\usepackage{titlesec}
\usepackage{amsmath}
\usepackage{url}
\usepackage{booktabs}
\usepackage{multirow}
\usepackage{enumitem}

\newcommand{\remove}[1]{}
\definecolor{Ora}{cmyk}{0, 0.6, 0.8, 0}

\begin{document}

\date{}

\title{\Large \bf RAPTOR: Routing Attacks on Privacy in Tor}

\author{
{\rm Yixin Sun}\\
Princeton University
\and
{\rm Anne Edmundson}\\
Princeton University
\and
{\rm Laurent Vanbever}\\
ETH Zurich
\and
{\rm Oscar Li}\\
Princeton University
\and
{\rm Jennifer Rexford}\\
Princeton University
\and
{\rm Mung Chiang}\\
Princeton University
\and
{\rm Prateek Mittal}\\
Princeton University
} %

\maketitle

\thispagestyle{empty}

\subsection*{Abstract}
The Tor network is a widely used system for anonymous communication. However, Tor is known to be vulnerable to attackers who can observe traffic at both ends of the communication path. In this paper, we show that prior attacks are just the tip of the iceberg. We present a suite of new attacks, called Raptor, that can be launched by Autonomous Systems (ASes) to compromise user anonymity.
First, AS-level adversaries can exploit the asymmetric nature of Internet routing to increase the chance of observing at least one direction of user traffic at both ends of the communication. Second, AS-level adversaries can exploit natural churn in Internet routing to lie on the BGP paths for more
users over time. Third, strategic adversaries can manipulate Internet routing via BGP hijacks (to discover the users using specific Tor guard nodes) and interceptions (to perform traffic analysis).
We demonstrate the feasibility of Raptor attacks by analyzing historical BGP data and Traceroute data as well as performing real-world attacks on the live Tor network, while ensuring that we do not harm real users. In addition, we outline the design of two monitoring frameworks to counter these attacks: BGP monitoring to detect control-plane attacks, and Traceroute monitoring to detect data-plane anomalies. Overall, our work motivates the design of anonymity systems that are aware of the dynamics of Internet routing.

\section{Introduction}

Anonymity systems aim to protect user identities from untrusted destinations and third parties on the Internet. 
Among all of them, the Tor network~\cite{dingledine:sec04} is the most widely used. As of February 2015, the Tor 
network comprises of 7{,}000 relays or proxies which together carry terabytes of traffic every day~\cite{tor-metrics}. 
Tor serves millions of users and is often publicized by political dissidents, whistle-blowers, law-enforcement, 
intelligence agencies, journalists, businesses and ordinary citizens concerned about the privacy of their 
online communications~\cite{tor-users}.

Along with anonymity, Tor aims to provide low latency and, as such, does not obfuscate packet timings or sizes. 
Consequently, an adversary who is able to observe traffic on both segments of the Tor communication channel 
(\emph{i.e.}, between the server and the Tor network, and between the Tor network and the client) can correlate
packet sizes and packet timings to deanonymize Tor clients~\cite{shmatikov:esorics06,syverson:wpet01}. 

There are essentially two ways for an adversary to gain visibility into Tor traffic, either by compromising (or owning enough) 
Tor relays or by manipulating the underlying network communications so as to put herself on the forwarding path 
for Tor traffic. Regarding network threats, large Autonomous Systems (ASes) such as Internet Service Providers (ISPs) can easily 
eavesdrop on a portion of all links, and observe any unencrypted information, packet headers, packet timing, and packet size.
Recent declarations by Edward Snowden have confirmed that ASes poses a real threat. Among others, the NSA has a program called 
Marina which stores meta information about user communications for up to a year~\cite{marina}, while the GCHQ has a program called
Tempora that stores meta-information for 30 days and buffers data for three days~\cite{mti}. Also, and maybe more importantly, 
it has been shown that Tor was targeted by such adversaries in collusion with ASes~\cite{quantum,torstinks,giraffe}. 

In this paper, we present Raptor, a suite of novel traffic analysis attacks that deanonymize Tor users more effectively
than previously thought possible. To do so, and unlike previous studies on AS-level adversaries~\cite{feamster:wpes04,edman:ccs09,murdoch:pet07},
Raptor leverages the \emph{dynamic aspects} of the Internet routing protocol, \emph{i.e.} the Border Gateway Protocol (BGP).

\begin{table*}[t]
\centering
\def\arraystretch{1.1}
  \begin{tabular}{@{}llclclclcl@{}}\toprule
   && Traffic Analysis && BGP Churn && BGP Hijack && BGP Interception \\
  \midrule
  Symmetric   && Known~\cite{shmatikov:esorics06,syverson:wpet01} && \multirow{2}{*}{\emph{Novel} (\S\ref{sec:churn})} && \multirow{2}{*}{\emph{Novel} (\S\ref{sec:hijacks_and_interception})} && \multirow{2}{*}{\emph{Novel} (\S\ref{sec:hijacks_and_interception})} \\
  Asymmetric  && \emph{Novel} (\S\ref{sec:asymmetry}) && && && \\
  \bottomrule
  \end{tabular}
\caption{This paper describes Raptor, a suite of previously unknown attacks on the Tor Network}
\label{tbl:tbl1}
\end{table*}

Raptor attacks are composed of three individual attacks whose effects are compounded (\S\ref{sec:raptor}). First, Raptor exploits the asymmetric nature of Internet routing: the BGP path from a sender to a receiver can be different than the BGP path from the receiver to the sender. Internet routing asymmetry increases the chance of an AS-level adversary observing at least one direction of both communication endpoints, enabling a novel asymmetric traffic analysis attack. Second, Raptor exploits natural churn in Internet routing: BGP paths change over time due to link or router failures, setup of  new Internet links or peering relationships, or changes in AS routing policies.  Changes in BGP paths allow ASes to observe additional Tor traffic, enabling them to deanonymize an increasing number of Tor clients over time. Third, Raptor exploits the inherent insecurity of Internet routing: strategic adversaries can manipulate Internet routing via BGP hijack and BGP interception attacks against the Tor network. These attacks enable the adversary to observe user communications, and to deanonymize clients via traffic analysis.

Raptor attacks were briefly discussed in a preliminary and short workshop paper~\cite{Vanbever:2014:AQU:2670518.2673869}. In this paper, we go further by measuring the importance of the attacks using real-world Internet control- and data-plane data. We also demonstrate the attacks feasibility by performing them on the live Tor network---\emph{with success}. No real Tor users were harmed in our experiments (\S\ref{sec:ethics}). Finally, we also describe efficient countermeasures to restore a good level of anonymity. To summarize, we make the following key contributions: 

\noindent\textbf{Asymmetric Traffic Analysis and BGP Churn:} Using live experiments on the Tor network, we showed that Raptor's asymmetric traffic analysis attacks can deanonymize a user with a 95\% accuracy, without any false positives (\S\ref{sec:asymmetry}). Using historical BGP and Traceroute data, we showed that by considering routing asymmetry and routing churn, the threat of AS-level attacks increases by 50\% and 100\%, respectively (\S\ref{sec:churn}).

\noindent\textbf{BGP Hijacks and Interceptions:} We analyzed known BGP hijacks and interception attacks on the Internet and show multiple 
instances where Tor relays were among the target prefixes (\S\ref{sec:hijacks_and_interception}). As an illustration, the recent Bitcoin Hijack attack~\cite{bitcoin} in 2014, 
as well as Indosat Hijack attacks~\cite{indosat2014,indosat} in 2014 and 2011 involved multiple 
Tor relays. To demonstrate the feasibility of such attacks for the purpose of deanonymizing Tor clients, we successfully performed an interception attack against a live Tor relay. Overall, we found that more than 90\% of Tor relays are vulnerable to our attacks. 

\noindent\textbf{Countermeasures:} We present a comprehensive taxonomy of countermeasures against Raptor attacks (\S\ref{sec:countermeasures}). In particular, we outline the design of 
a monitoring framework for the Tor network that aims to detect suspicious AS-level path changes towards Tor prefixes using both BGP and Traceroute monitoring.

\section{Raptor Attacks}
\label{sec:raptor}

To communicate with a destination, Tor clients establish layered circuits through three subsequent Tor relays. 
The three relays are referred to as: \emph{entry} (or \emph{guard}) for the first one, \emph{middle} for the second one,
and \emph{exit} relay for the last one. To load balance its traffic, Tor clients select relays with a probability that is proportional 
to their network capacity. Encryption is used to ensure that each relay learns the identity of only the previous hop and the next hop
in the communications, and no single relay can link the client to the destination server.

It is well known that if an attacker can observe the traffic from the destination server to the exit relay as well as from the entry relay to the 
client (or traffic from the client to the entry relay and from the exit relay to the destination server), then it can leverage correlation between packet timing and sizes to infer the network identities of clients and servers (end-to-end timing analysis). This timing analysis works even if the communication is encrypted. 

In the rest of the section, we present the three Raptor attacks and how they contrast to conventional symmetric 
traffic analysis. We start by discussing how seeing
just one direction of the traffic for each segment (between
the sender and the guard, and between the last relay
and the destination) is sufficient for the adversary (\S\ref{ssec:asymmetric}). We then
explain how ASes can exploit natural BGP dynamics (\S\ref{ssec:churn}), 
or even launch active attacks (\S\ref{ssec:hijack}), to compromise the anonymity of Tor users.

\subsection{Asymmetric Traffic Analysis}
\label{ssec:asymmetric}

\begin{figure}[h]
 \centering
 \begin{subfigure}[t]{0.49\columnwidth}
 \includegraphics[width=\columnwidth]{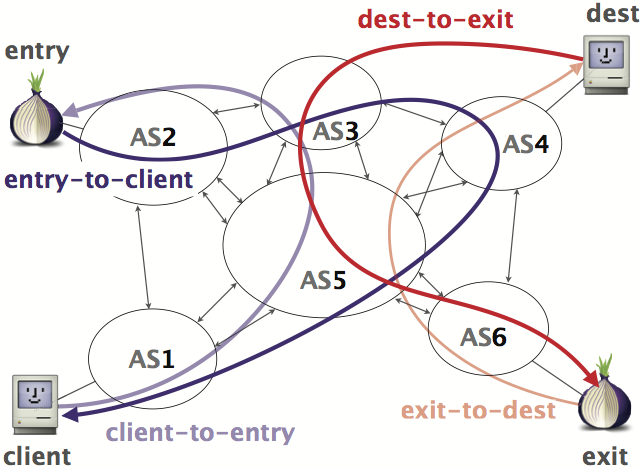}
 \label{fig:asymmetry_left}
 \end{subfigure}
 \begin{subfigure}[t]{0.49\columnwidth}
 \includegraphics[width=\columnwidth]{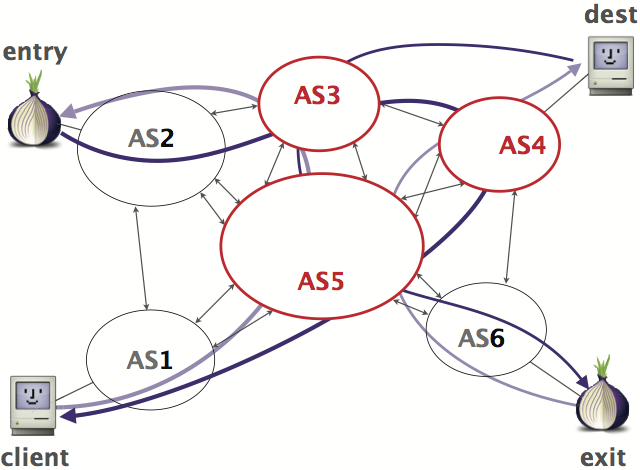}
 \label{fig:asymmetry_right}
 \end{subfigure}
 \caption{Asymmetric routing increases the power of AS-level adversaries. When considering forward traffic, i.e., client-to-entry and exit-to-destination flows, 
only AS5 can compromise anonymity. When considering both forward and backward traffic though, AS3, AS4 and AS5 can compromise anonymity. 
Our measurements confirm that asymmetric traffic analysis is feasible.}
 \label{fig:asymmetry}
\end{figure}

We propose asymmetric traffic analysis, a novel form of end-to-end timing analysis that allows AS-level adversaries to 
compromise the anonymity of Tor users. Let us suppose that a Tor client is uploading a large file to a Web server. 
Conventional traffic analysis considers only one scenario where adversaries observe traffic from the client to the entry 
relay, and from the exit relay to the Web server (same direction as the flow of traffic)\footnote{If the traffic 
is flowing from the server to the client, then end-to-end timing analysis considers a scenario
where the adversary observes traffic from the Web server to the exit relay and from the entry relay to the client.}.

However, Internet paths are often asymmetric: the path from the exit relay to the Web server may be different than the 
path from the Web server to the exit relay. Thus it is possible that an adversary may not be able to observe the data traffic 
on the path from the exit relay to the server, but it observes the TCP acknowledgment traffic on the path 
from the server to the exit relay. 

We introduce an asymmetric traffic analysis attack that allows an adversary to deanonymize users as long as the adversary is 
able to observe \emph{any direction of the traffic}, at both ends of the communication. Note that we can 
view the conventional end-to-end timing analysis as a special case of our attack, in which the adversary is 
able to observe traffic at both ends of the anonymous path, and in the same direction as the flow of traffic.
Routing asymmetry increases the number of ASes who can observe at least 
one direction of traffic at both communication endpoints. We illustrate this scenario in Figure~\ref{fig:asymmetry}.

More concretely, our attack is applicable to four scenarios where an adversary observes (a) data traffic from the client to entry 
relay, and data traffic from exit relay to the server, or (b) data traffic from the client to entry relay, and TCP acknowledgment 
traffic from the server to exit relay, or (c) TCP acknowledgment traffic from guard relay to the client, and data traffic from exit 
relay to the server, or (d) TCP acknowledgment traffic from guard relay to the client, and TCP acknowledgment traffic from 
the server to the exit relay.

A key hurdle in asymmetric traffic correlation is that TCP acknowledgments are cumulative, and there is
not a one-to-one correspondence between data packets and the TCP acknowledgment packets. We overcome this hurdle 
by observing that Tor (and other anonymity systems) use SSL/TLS for encryption, which leaves the TCP header unencrypted. 
Our attack inspects TCP headers in the observed traffic to retrieve the TCP sequence number field and TCP 
acknowledgment number field, and analyzes the correlation between these fields of both ends over time. 
Our experimental results in Section~\ref{sec:asymmetry} show the feasibility of asymmetric traffic analysis, 
with a detection accuracy of 95\%. Furthermore, asymmetric traffic analysis can be combined with other Raptor 
attacks, such as exploiting natural churn and BGP interception attack, which we discuss next.

\subsection{Natural Churn}\label{ssec:churn}

\begin{figure}[h]
 \centering
 \begin{subfigure}[t]{0.49\columnwidth}
 \includegraphics[width=\columnwidth]{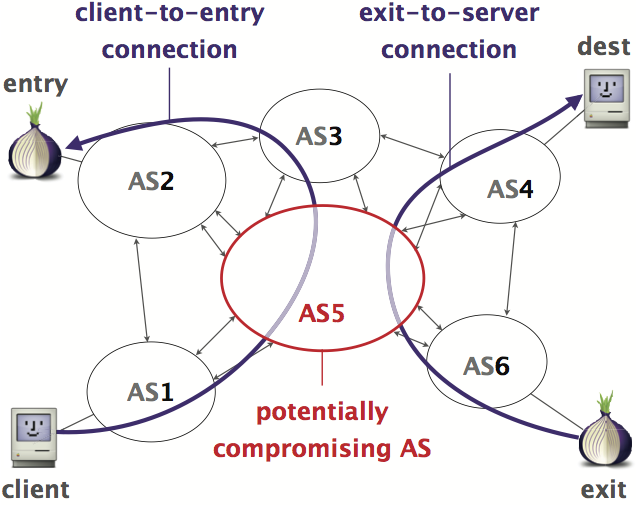}
 \label{fig:churn_l}
 \end{subfigure}
 \begin{subfigure}[t]{0.49\columnwidth}
 \includegraphics[width=\columnwidth]{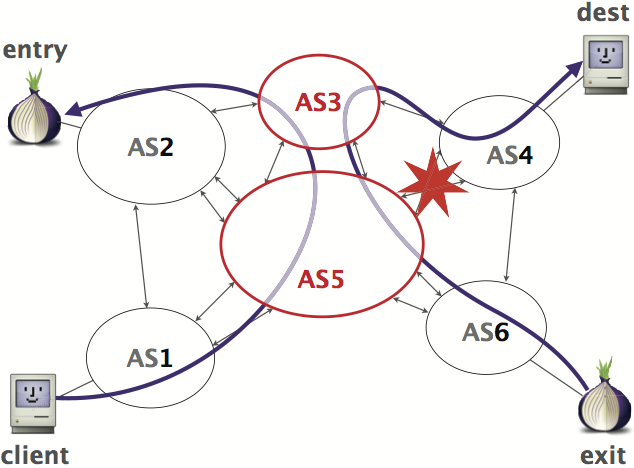}
 \label{fig:churn_r}
 \end{subfigure}
 \caption{BGP churn increases the number of ASes that can deanonymize Tor traffic. Initially, only AS5 can deanonymize the client, seeing both direction of the traffic (left). After the failure of link (AS4, AS5), both AS5 and AS3 can deanonymize Tor traffic (right).}
 \label{fig:churn_example}
\end{figure}

When users communicate with recipients over multiple time instances, then there is a 
potential for compromise of anonymity at every communication instance~\cite{wright:sp03,overlier:sp06}.
Thus anonymity can degrade over time. Tor considers this threat from the perspective of adversarial 
relays (but not adversarial ASes).%
 Tor clients use a fixed entry relay (guard relay) for a period of time (Dingledine et al. recommend 9 months~\cite{dingledine:hotpets14}) to mitigate 
this threat with respect to adversarial relays. We note that the threat of AS-level adversaries still persists, because even though the 
entry relay is fixed, the set of ASes on the path between the client and the guard relay may change over time. Next, we discuss such 
attacks that rely on natural churn in BGP paths.  

The underlying Internet paths between a client and guard relay vary over time due to changes
in the physical topology (\emph{e.g.}, failures, recoveries, and the rollout
of new routers and links) and AS-level routing policies (\emph{e.g.}, traffic
engineering and new business relationships).  These changes give a
malicious AS surveillance power that increases over time. For
example, AS 3 in Figure~\ref{fig:churn_example} does not lie on the
original path from the exit to the destination, but a BGP routing change
can put AS 3 on the path, allowing it to perform traffic analysis.

In Section~\ref{sec:churn}, we show that the surveillance capability of an AS-level 
adversary can increase up to 50\% when considering BGP churn over a period of one month.

\subsection{BGP Hijack}\label{ssec:hijack}

So far, we discussed Raptor attacks that were \emph{passive}. Strategic AS-level adversaries 
are also capable of launching \emph{active} attacks, that deviate from honest routing behavior. 
Internet routing is vulnerable to attacks which enable an AS to manipulate inter-domain routing 
by advertising incorrect BGP control messages. While these attacks are well known in the 
networking community, we are the first to apply these attacks to anonymity systems such 
as Tor.

AS-level adversaries can hijack an IP prefix~\cite{zhang:conext07} by advertising the prefix as 
its own. The attack causes a fraction of Internet traffic destined to the prefix to be captured 
by the adversary. Tor relay nodes can observe a large amount of client traffic. For example, a 
Tor guard relay observes information about client IP addresses. Thus, the IP prefixes corresponding 
to Tor guard relays presents an attractive target for BGP hijack.

As a concrete attack example, we consider a scenario where an AS-level adversary aims to deanonymize the 
user associated with a connection to a sensitive Web server (say a whistleblowing website). The adversary 
can first use existing attacks on the Tor network to uncover the identity of the client's guard 
relay~\cite{murdoch:sp05,mittal:ccs11,hopper:tissec10,overlier:sp06}. Next, the adversary can launch 
a BGP hijack attack against the Tor relay. This allows the adversary to see traffic destined to the 
guard relay. 
BGP hijack thus enables an adversary to learn the set of all client IP addresses (anonymity set) associated with a guard relay (and the 
target connection to the sensitive Web server). 

We note that in a prefix-hijack attack, the captured traffic is blackholed, and the client's 
connection to the guard is eventually dropped. Thus, it may not be possible to perform fine-grained traffic 
analysis to infer the true client identity from this anonymity set. However, the identification of 
a reduced anonymity set (as opposed to the entire set of Tor users) is already a significant amount 
of information leakage, and can be combined with other contextual information to break user 
anonymity~\cite{tor-harvard}. In Section~\ref{sec:hijacks_and_interception}, we uncover several real-world BGP hijack 
attacks in which Tor relays were among the target prefixes. 

\subsection{BGP Interception}

\begin{figure}[h]
 \centering
 \begin{subfigure}[t]{0.49\columnwidth}
 \includegraphics[width=\columnwidth]{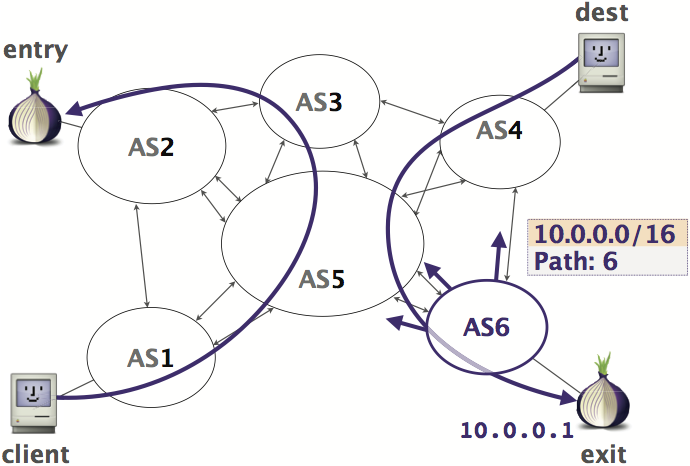}
 \label{fig:hijack_left}
 \end{subfigure}
 \begin{subfigure}[t]{0.49\columnwidth}
 \includegraphics[width=\columnwidth]{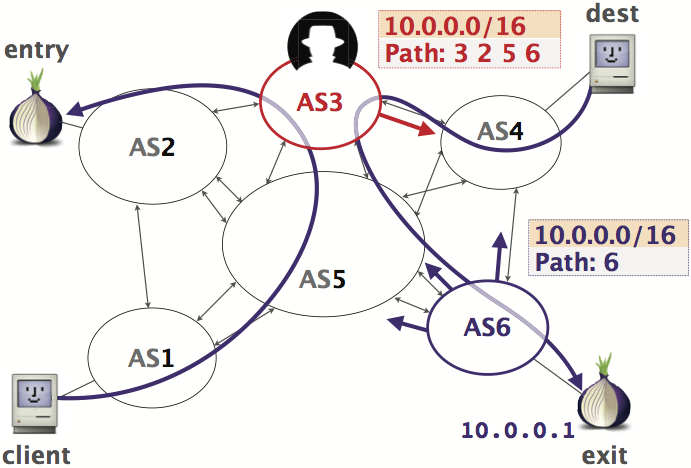}
 \label{fig:hijack_right}
 \end{subfigure}
 \caption{BGP interception attack enables ASes to selectively put themselves on some path. Here, AS3 only sees 
traffic between the client and the entry relay (left). By intercepting the prefix containing the exit relay (right), 
AS3 also sees traffic towards the exit relay, enabling it to deanonymize the Tor communication.}
 \label{fig:interception}
\end{figure}

Our BGP hijack attack discussed above allows adversaries to capture traffic destined towards a target Tor prefix, 
but the captured traffic is blackholed, resulting in the connecting being dropped. Next, we discuss a more sophisticated 
routing attack called BGP interception attack~\cite{ballani:sigcomm07}, that allows adversaries to perform exact deanonymization of Tor users.

A prefix interception attack allows the malicious AS to become an intermediate AS in the path towards the 
guard relay, \emph{i.e.}, after interception, the traffic is routed back to the actual guard relay. Such an 
interception attack allows the connection to be kept alive, enabling the malicious AS to exactly deanonymize 
the client via asymmetric traffic analysis.

Similar to the previous discussion, let us consider an adversary trying to deanonymize the user 
connecting to a sensitive website (the adversary already sees the traffic towards the website). The adversary can first uncover the identity of the 
guard relay using existing attacks~\cite{murdoch:sp05} (as before),
and then launch a prefix interception attack against the guard relay. Since the adversary 
routes the traffic back to the guard relay, the client's connection is kept alive, allowing 
the adversary to launch asymmetric traffic correlation attacks. Note that in contrast to 
BGP hijack attacks, BGP interception attacks can perform exact deanonymization of Tor clients.  

These attacks enable malicious ASes to deanonymize user identity corresponding 
to a monitored target connection. Similarly, ASes that already see the client's traffic 
to its guard can position themselves to observe the traffic between the server and the 
exit relay by launching interception attacks against exit relays. Figure~\ref{fig:interception} 
illustrates this attack scenario.  

Finally, we note that a remote adversary can launch interception attacks against both guard relays 
and exit relays simultaneously, to perform general surveillance of the Tor network. In 
Section~\ref{sec:hijacks_and_interception}, we demonstrate a real-world BGP interception 
attack against a live Tor relay by collaborating with autonomous system operators.

\section{Asymmetric Traffic Analysis}\label{sec:asymmetry}
In this section, we experimentally show that asymmetric traffic analysis 
attacks are feasible. We use the live Tor network for our experiments. To 
protect the safety of real Tor users, we generate our own traffic through the 
Tor network. Our goal is to investigate the accuracy of asymmetric traffic 
analysis in deanonymizing our generated traffic.

\noindent{\bf Experimental Setup:} In order to generate our own traffic through the live Tor network, we use PlanetLab nodes 
as clients and Web servers. PlanetLab is an open platform for networking 
research, that provides access to hundreds of geographically distributed machines. We randomly 
pick $100$ machines on PlanetLab, located across United States, Europe, and Asia. 
We installed Tor clients on $50$ of those machines, and used the Privoxy tool 
(\url{www.privoxy.org}) to configure wget requests to tunnel over Tor. The remaining 
$50$ machines were setup to be Web servers, each containing a $100MB$ image file. 

We use the default Tor configuration on the $50$ client machines. We launch wget requests on 
the $50$ clients at the same time, each requesting a $100$MB image file from one of the $50$ web 
servers, respectively. 
We use tcpdump to capture data for $300$ seconds at the clients and the servers during this process.

\noindent{\bf Asymmetric correlation analysis:}
In each packet trace, we first extract the TCP sequence number and TCP acknowledgment number fields 
in the TCP header. 
Using the TCP sequence and acknowledgment numbers, we next compute the number of transmitted data 
bytes per unit time. For each pair of observed traces, we compute the correlation between 
the vector of transmitted data bytes over time. For our analysis, we use the Spearman's rank correlation 
coefficient (other correlation metrics could also be applicable). For each client, our 
asymmetric traffic analysis attack selects the server trace with the highest correlation as the 
best match.

\begin{figure}[t]
\begin{subfigure}{.25\textwidth}
  \centering
  \includegraphics[width=.9\linewidth]{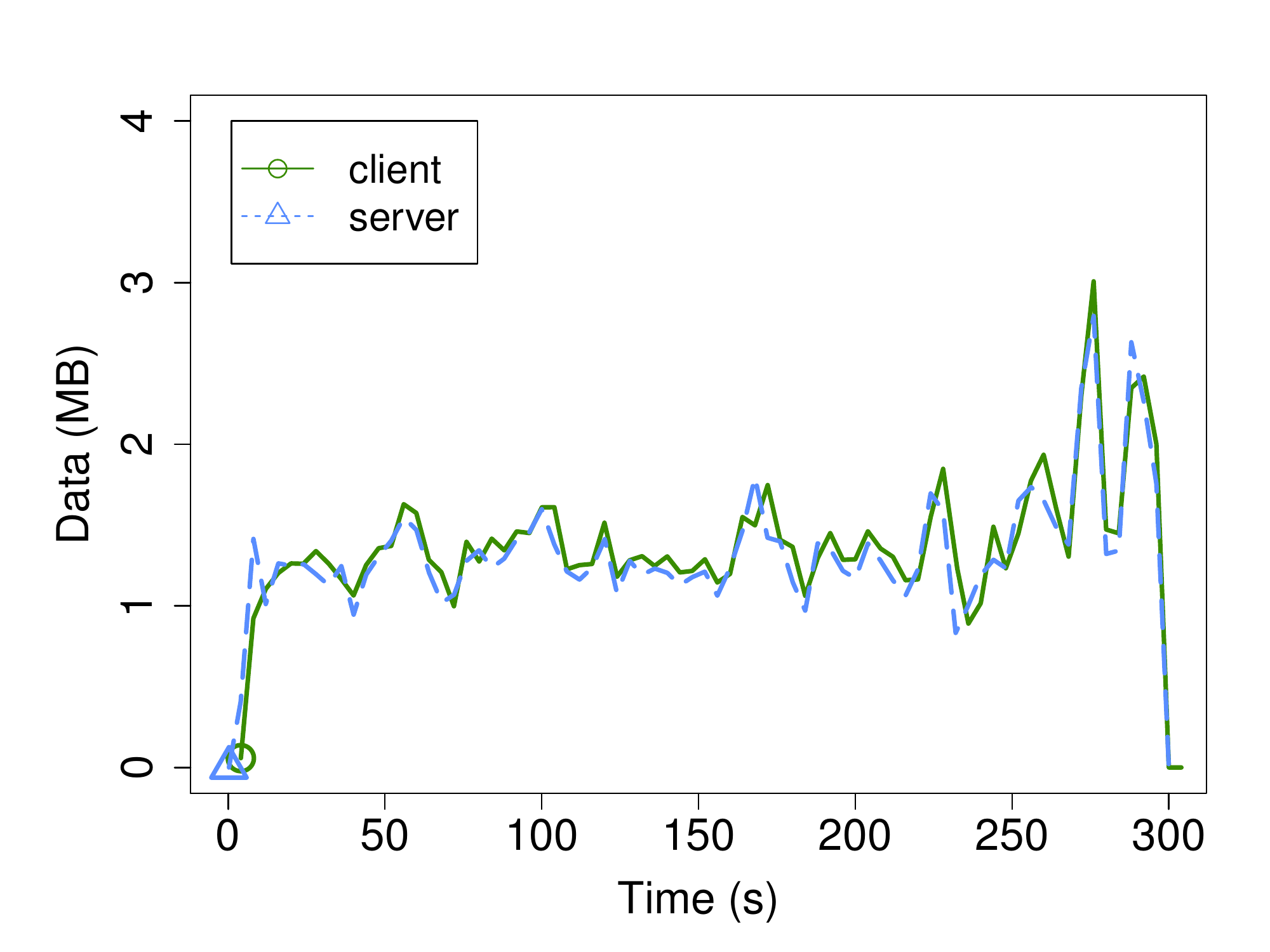}
  \caption{Client: ACK, Server: ACK}
\end{subfigure}%
\begin{subfigure}{.25\textwidth}
  \centering
  \includegraphics[width=.9\linewidth]{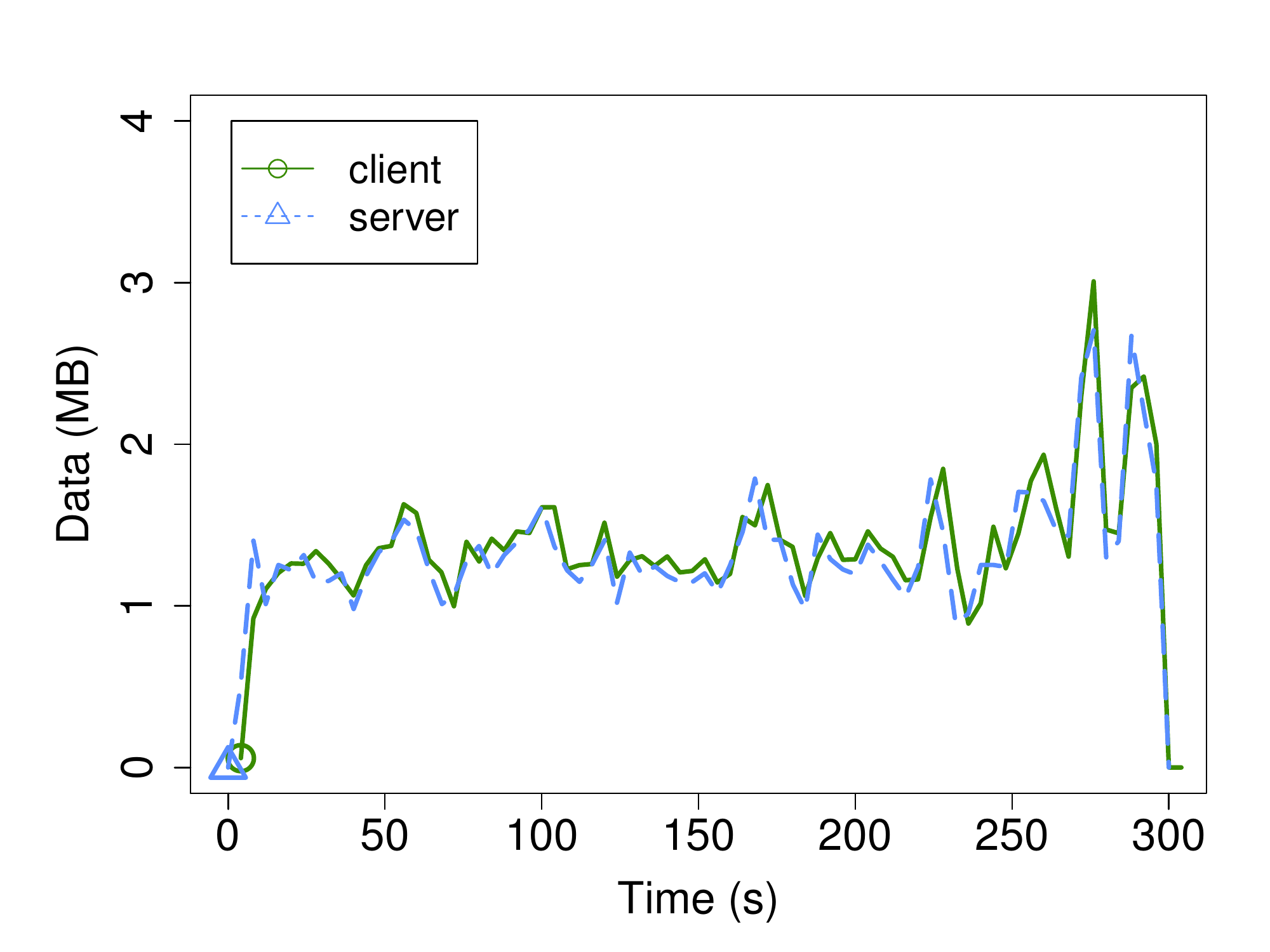}
  \caption{Client: ACK, Server: Data}
\end{subfigure}\\
\begin{subfigure}{.25\textwidth}
  \centering
  \includegraphics[width=.9\linewidth]{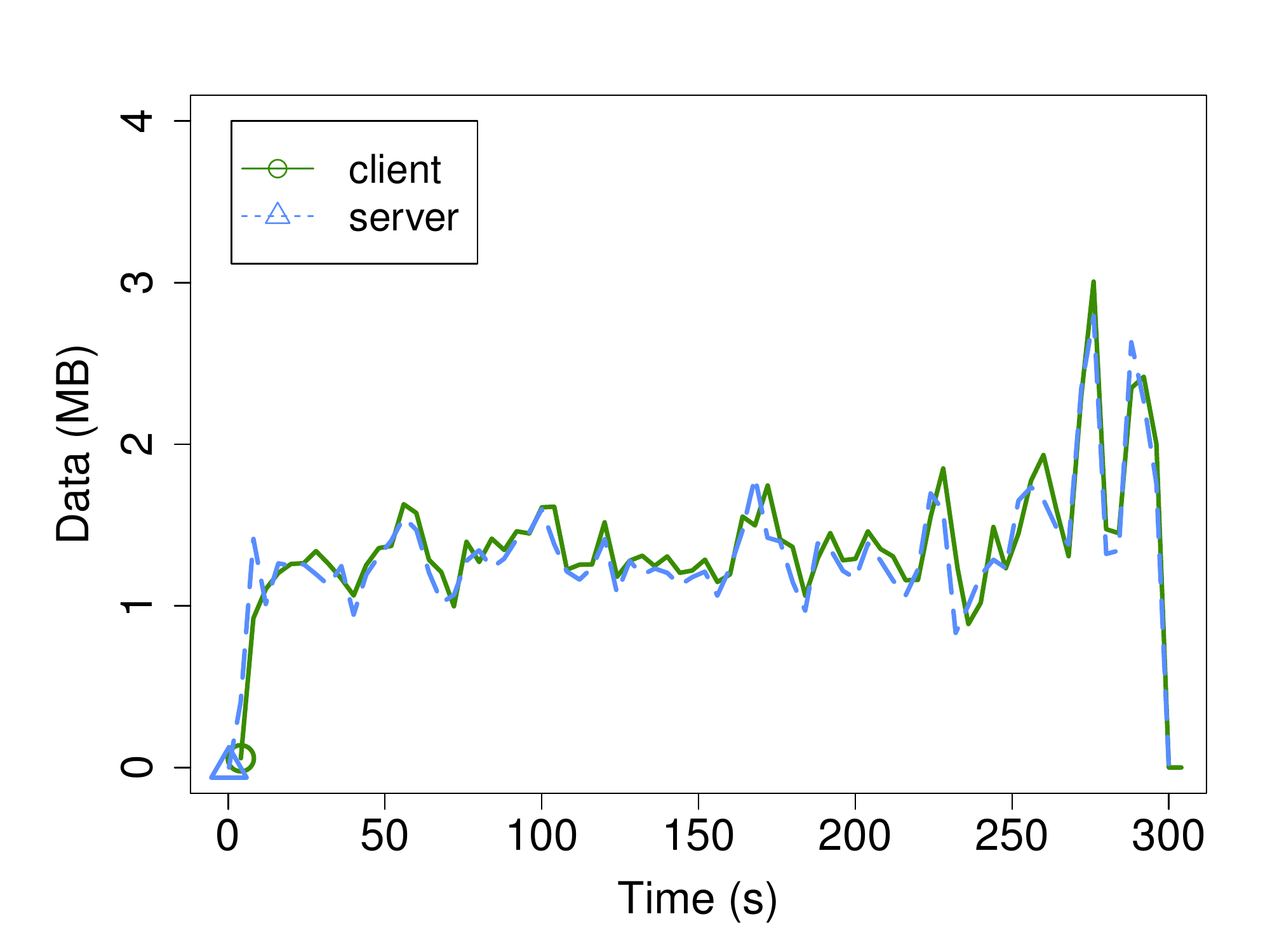}
  \caption{Client: Data, Server: ACK}
\end{subfigure}%
\begin{subfigure}{.25\textwidth}
  \centering
  \includegraphics[width=.9\linewidth]{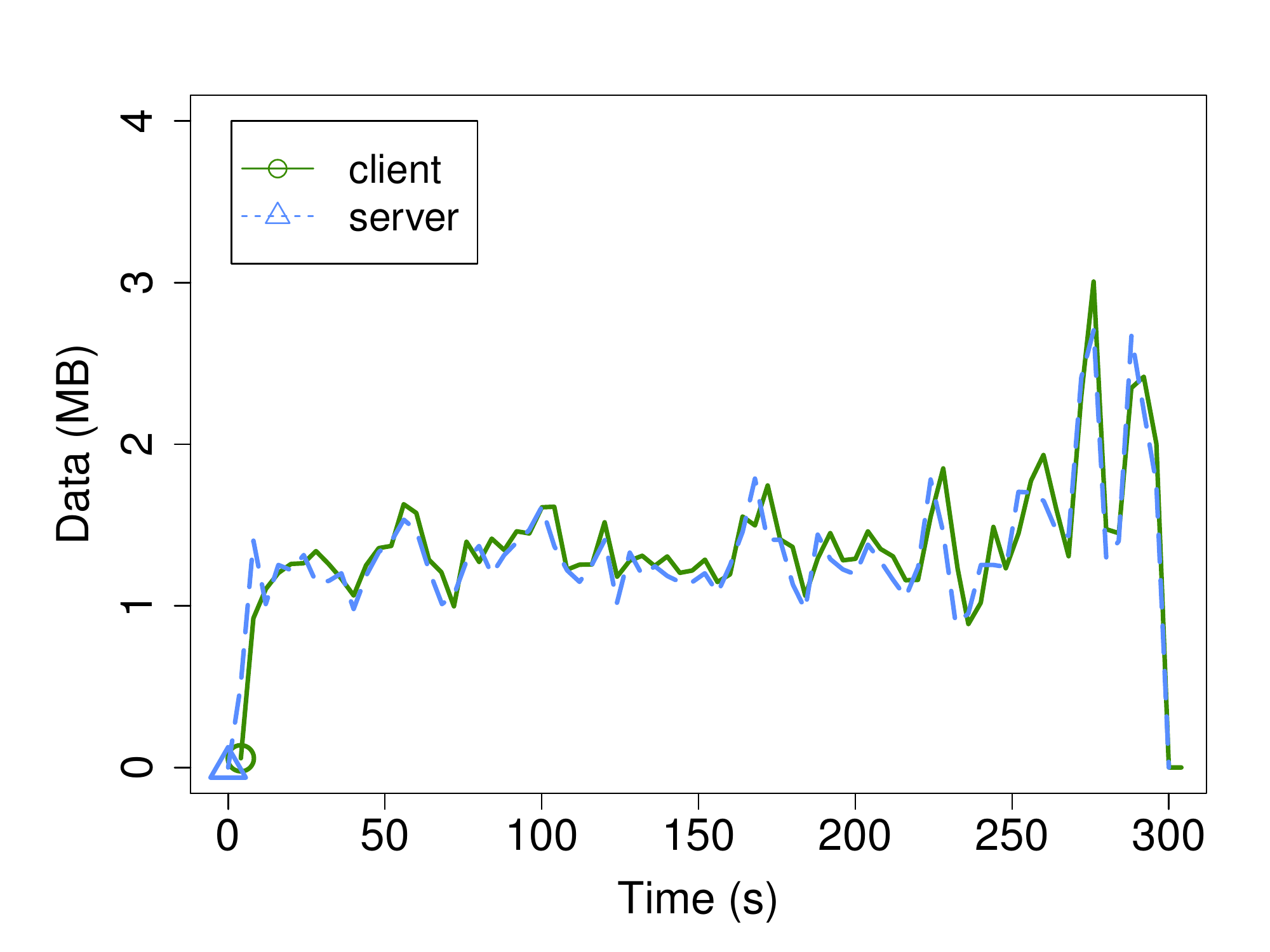}
  \caption{Client: Data, Server: Data}
\end{subfigure}\\
\caption{Asymmetric traffic analysis shows high correlation between a matched client/server pair}
\label{fig:asym_corr}
\end{figure}

\begin{figure}[t]
\begin{subfigure}{.25\textwidth}
  \centering
  \includegraphics[width=.9\linewidth]{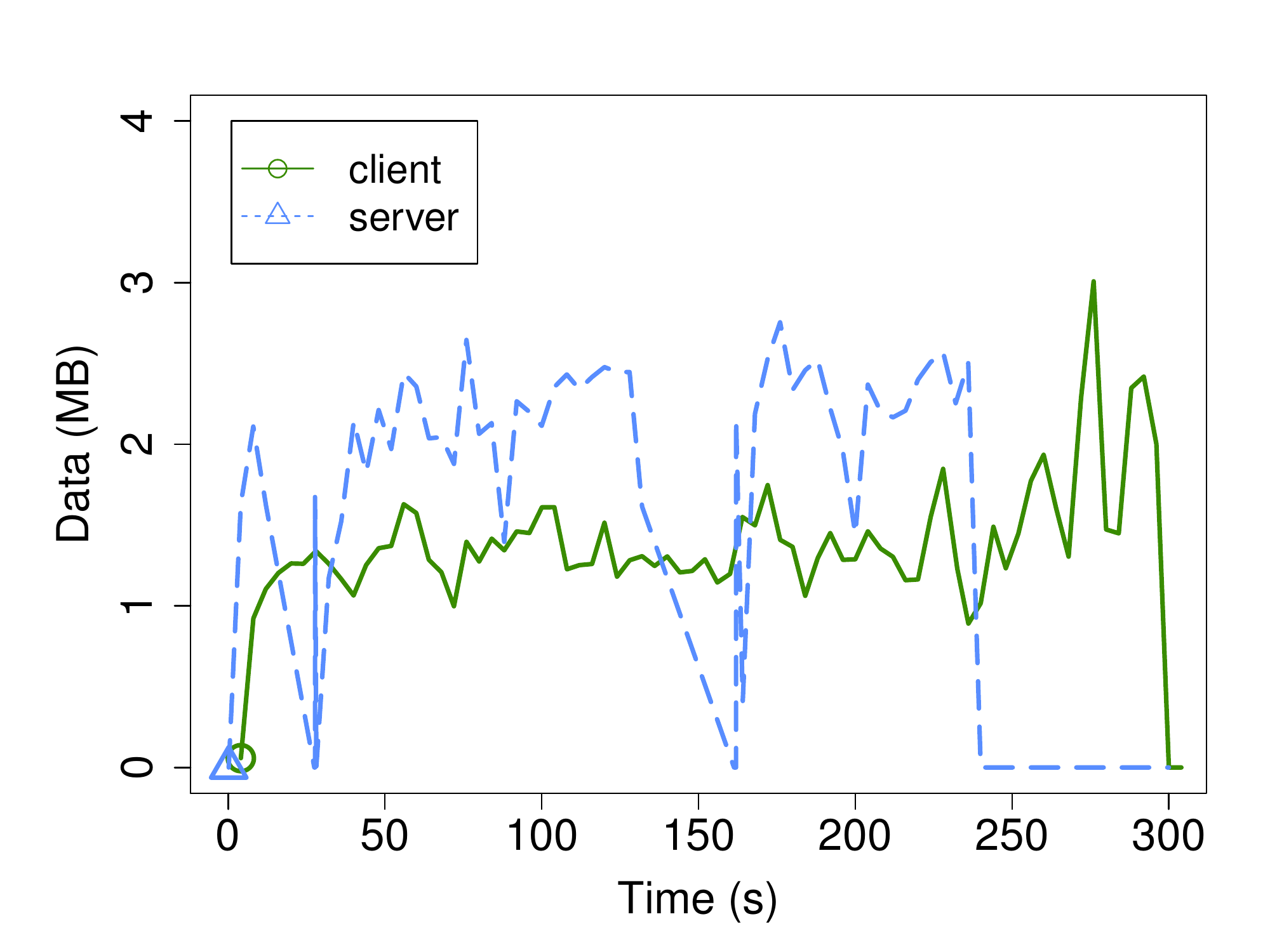}
  \caption{Client: ACK, Server: ACK}
\end{subfigure}%
\begin{subfigure}{.25\textwidth}
  \centering
  \includegraphics[width=.9\linewidth]{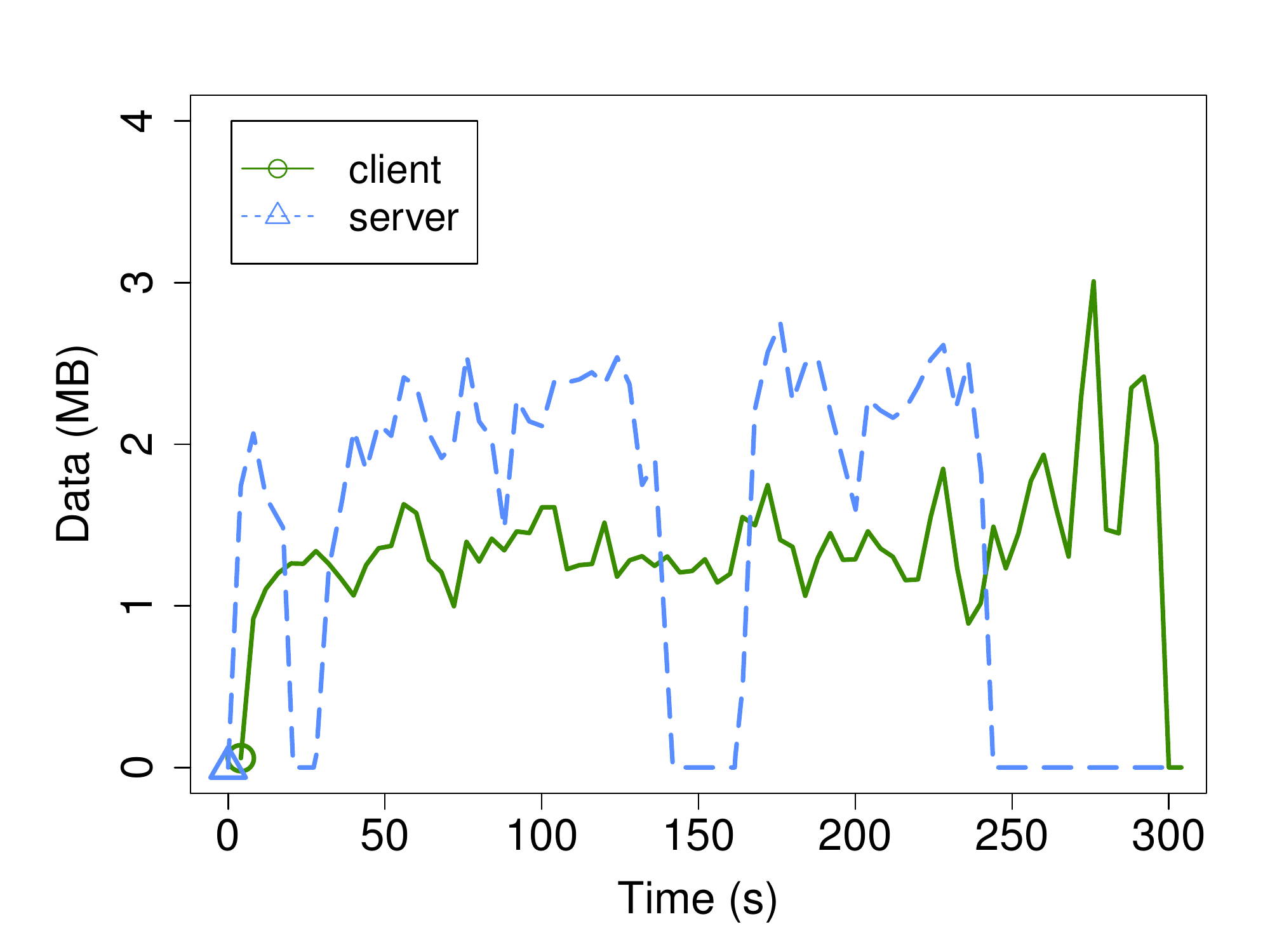}
  \caption{Client: ACK, Server: Data}
\end{subfigure}\\
\begin{subfigure}{.25\textwidth}
  \centering
  \includegraphics[width=.9\linewidth]{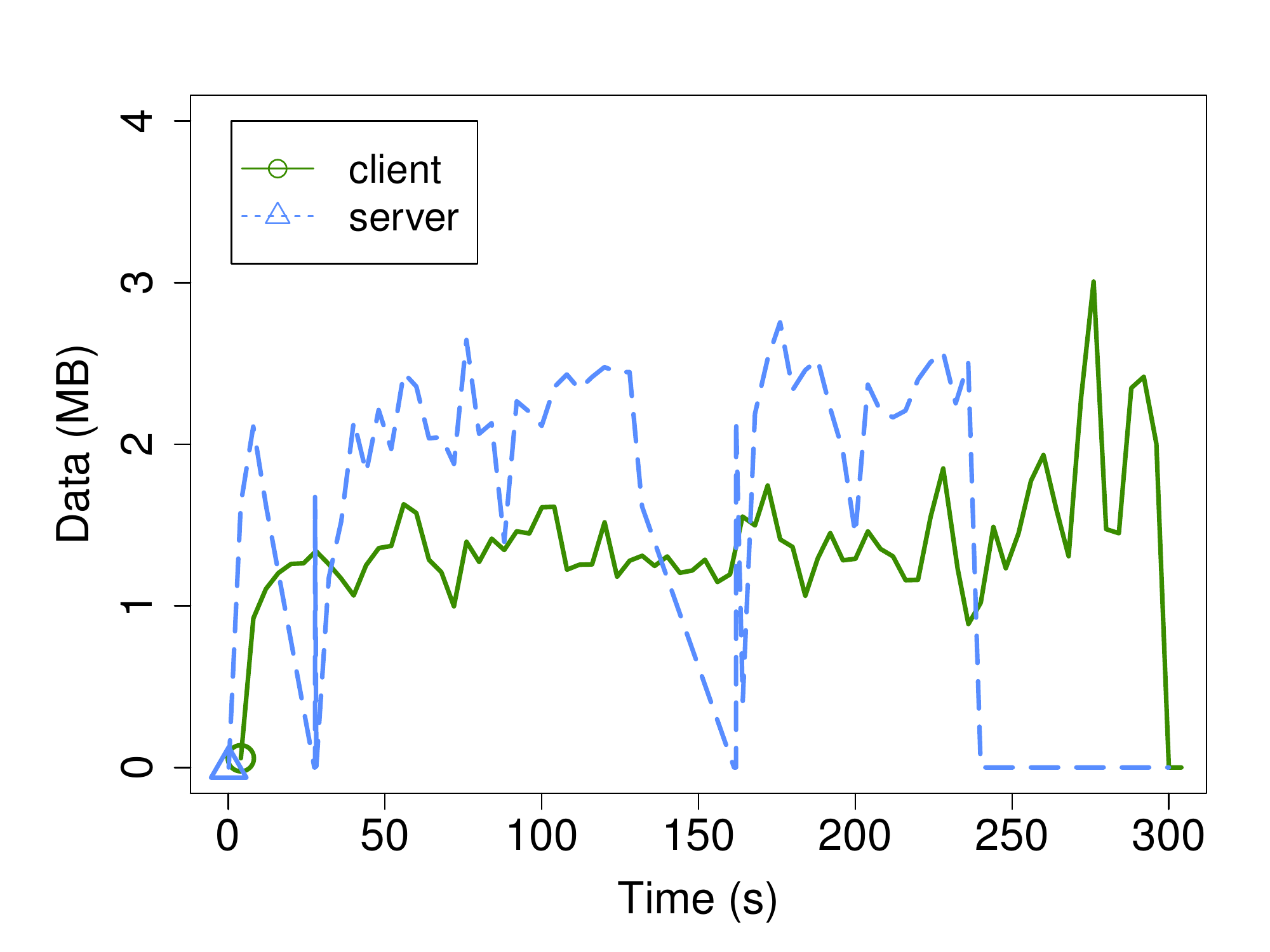}
  \caption{Client: Data, Server: ACK}
\end{subfigure}%
\begin{subfigure}{.25\textwidth}
  \centering
  \includegraphics[width=.9\linewidth]{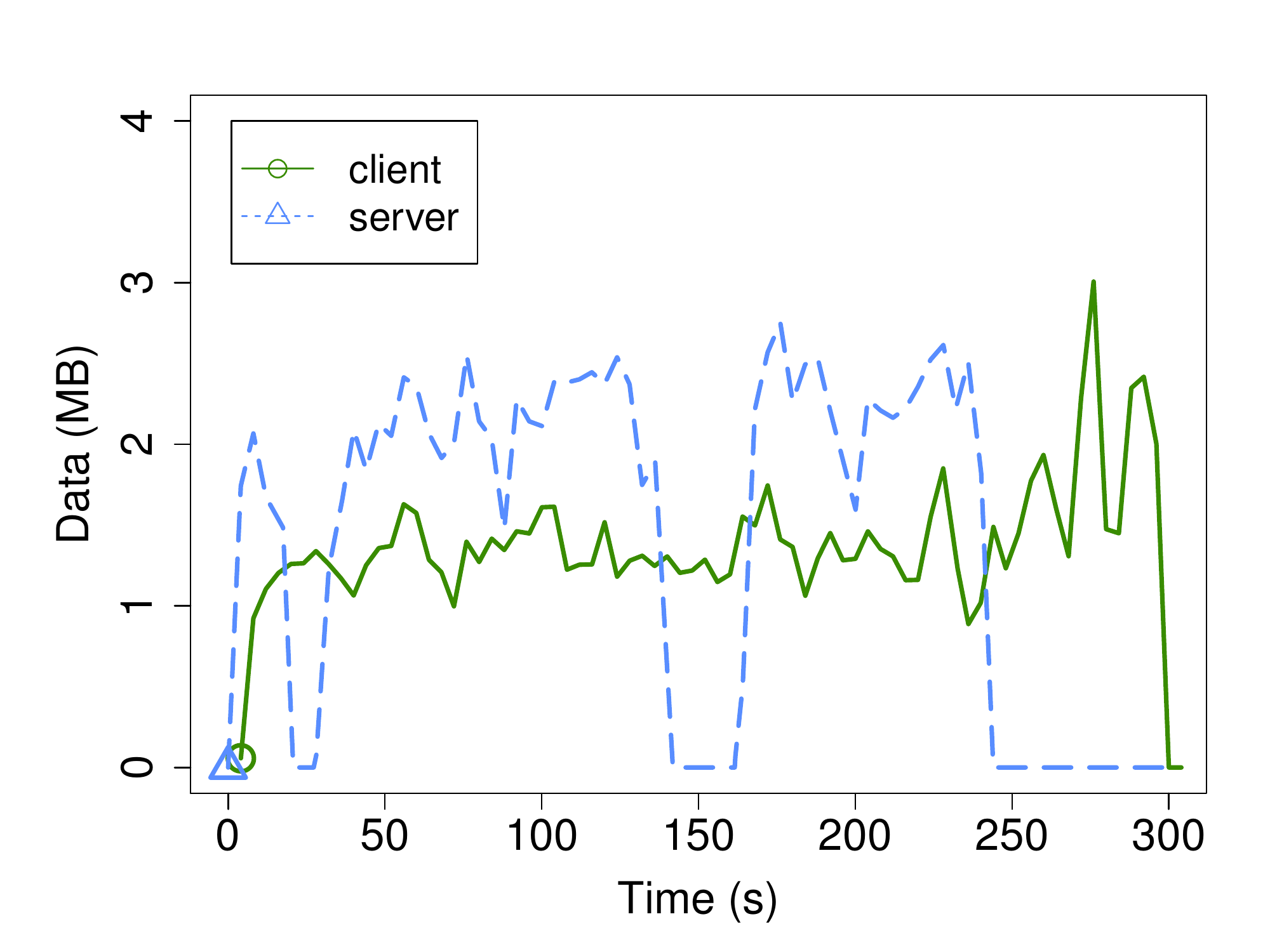}
  \caption{Client: Data, Server: Data}
\end{subfigure}\\
\caption{Asymmetric traffic analysis shows low correlation between an unmatched client/server pair}
\label{fig:asym_uncorr}
\end{figure}

\begin{figure}
\centering
\includegraphics[width=.6\linewidth]{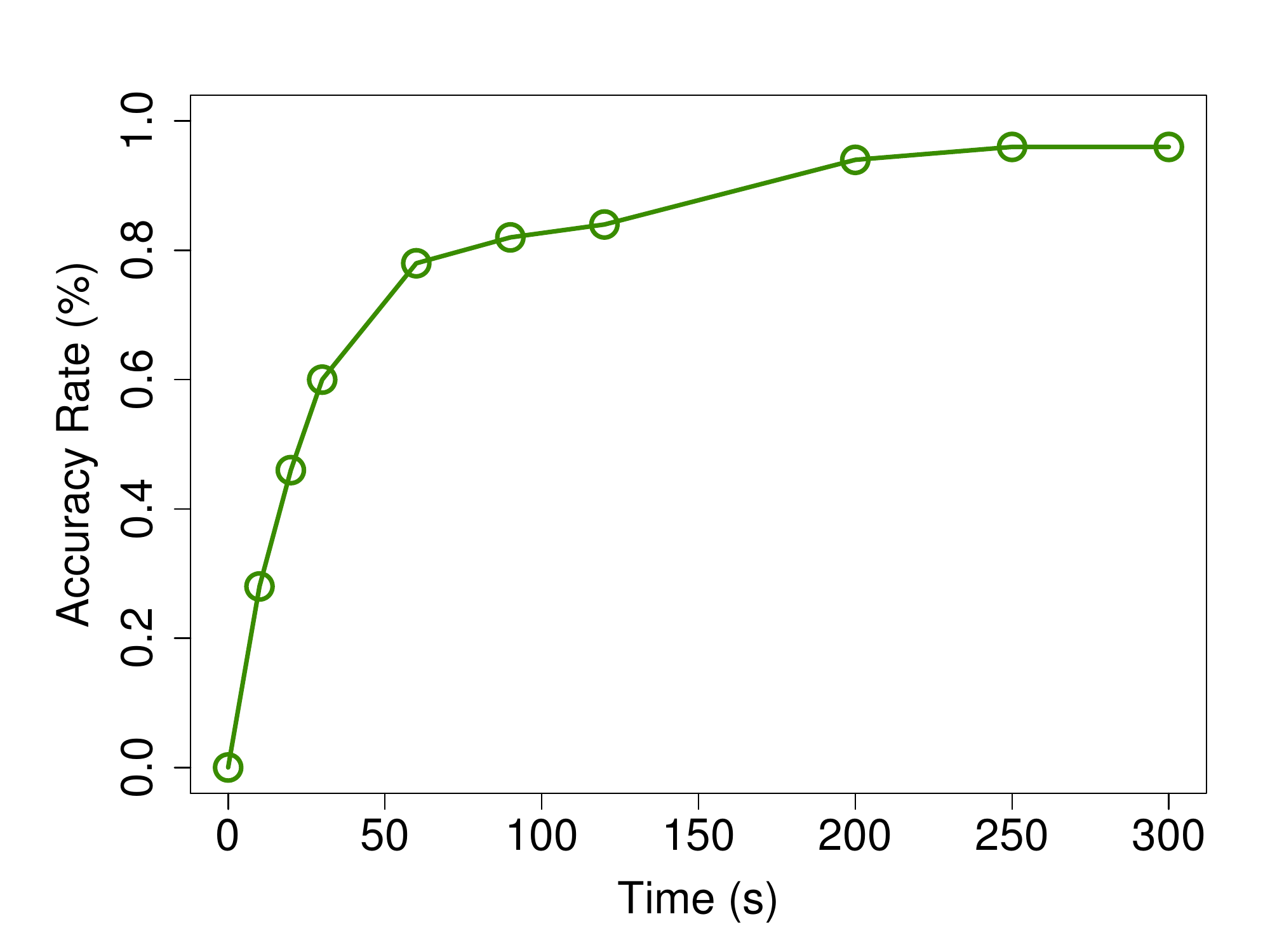}
\caption{The accuracy of the attack quickly 
increases with time, reaching 80\% within a minute, 95\% after five minutes.}
\label{fig:accu_attack}
\end{figure}

\noindent{\bf Results:} Figure~\ref{fig:asym_corr} illustrates our asymmetric analysis computed 
between a client server pair that is communicating. We can see high correlation in 
all four observation scenarios discussed in Section 2. Figure~\ref{fig:asym_uncorr} illustrates our 
asymmetric analysis computed between a client server pair that is not communicating 
with each other. 
We can see that incorrectly matched pairs have poor correlation in all four observation 
scenarios. Figure ~\ref{fig:accu_attack} illustrates the detection accuracy rate grows as the duration of attack increases, 
especially in the first $30$ seconds. 

We computed the detection accuracy of our asymmetric traffic analysis 
attacks in all four scenarios after $300$ seconds (by selecting the highest correlated pair), 
and obtained an average accuracy of $95\%$ (Table~\ref{tbl:planetlab-mean}). The error matches are 
all false negatives, for which the client has insignificant correlation coefficients with all servers, 
so it fails to be matched to any servers. We did not observe any false positives in our results.

\begin{table}[ht]
\scriptsize
\centering
\def\arraystretch{1.1}
  \begin{tabular}{@{}lcccc@{}}\toprule
  & Client ACK/ & Client ACK/ & Client Data/ & Client Data/ \\
  & Server ACK & Server Data & Server ACK & Server Data \\
  \midrule
  Overall         & 96\% & 94\% & 96\% & 94\% \\
  False negative  & 4\% & 6\% & 4\% & 6\% \\
  False positive  & 0\% & 0\% & 0\% & 0\% \\
  \bottomrule
  \end{tabular}
\caption{Asymmetric traffic analysis accuracy rate}
\label{tbl:planetlab-mean}
\end{table}

In addition to the actual observed error rate above, we also performed a statistical tests  to 
compute the $95\%$ confidence interval on our error rate, given our sample size of $50$ client machines 
and $50$ server machines. Table~\ref{tbl:planetlab-bounds} illustrates the confidence intervals on our error rates.

\begin{table}[ht]
\scriptsize
\centering
\def\arraystretch{1.1}
  \begin{tabular}{@{}lllll@{}}\toprule
  & Client ACK/ & Client ACK/ & Client Data/ & Client Data/ \\
  & Server ACK & Server Data & Server ACK & Server Data \\
  \midrule
  \multirow{2}{*}{False negative}  & 0.48\%~-- & 1.25\%~-- & 0.48\%~-- & 1.25\%~-- \\
  & 13.71\% & 16.54\% & 13.71\% & 16.54\% \\
  \multirow{2}{*}{False positive}  & 0\%~-- & 0\%~-- & 0\%~-- & 0\%~-- \\
  & 0.15\% & 0.15\% & 0.15\% & 0.15\% \\
  \bottomrule
  \end{tabular}
\caption{Asymmetric traffic analysis error rate confidence interval}
\label{tbl:planetlab-bounds}
\end{table}

\section{Natural Churn}
\label{sec:churn}

In this section, we study and evaluate how routing dynamics, or churn, increase the power of AS-level adversaries in anonymity systems such as Tor. We start with an exhaustive control-plane analysis using collected BGP data (\S\ref{ssec:cp_churn_eval}). Our results show that churn can increase the amount of compromised Tor circuits by up to 50\% over a period of one month. We then confirmed our results by performing targeted data-plane measurements on the Tor network (\S\ref{ssec:dp_churn_eval}). Again, churn significantly increased the percentage of vulnerable Tor circuits, nearly tripling it.

\subsection{Control-plane Evaluation}
\label{ssec:cp_churn_eval}

We quantified the impact of churn by measuring how it increased the probability of a single AS (say AS $X$) to end 
up simultaneously on the path between a client and a guard relay and on the path between a destination and an exit relay. When this happens, we considered AS $X$ as (potentially) compromising for the pair (client, destination) using the corresponding Tor circuit. Observe that our analysis leverages asymmetric traffic analysis (\S\ref{sec:asymmetry}) as it only requires $X$ to be on-path for two publicly-known prefixes, covering the guard and the exit relay.

\noindent\textbf{Datasets} We collected 612+ \emph{million} BGP updates pertaining to 550{,}000 IP prefixes 
collected by six RIPE-maintained BGP Looking Glass ($rrc00$, $rrc01$, $rrc03$, $rrc04$, $rrc11$, $rrc14$)~\cite{ripe:ris} 
in January 2015 over 250+ BGP sessions. We processed the dataset to remove any artifacts caused by 
session resets~\cite{cheng2010longitudinal}. In parallel, we also collected Tor-related data 
(IP address, flags and bandwidth) of about 6755 Tor relays active during the 
same period of time~\cite{CollecTor}. Among all Tor relays, 1459 (resp. 1182) of them were listed 
as guards (resp. exits) and 338 relays were listed as both guard and exit.

We considered each BGP session as a proxy for Tor clients and destinations.
Note that analysis implicitly accounts for \emph{any} Internet host reachable directly or indirectly through 
these BGP sessions. %
Our dataset contains sessions belonging to major Internet transit providers such as Level-3, ATT, NTT, etc. 
that provide transit to millions of hosts.

\noindent\textbf{Static baseline.} We computed a static baseline by considering the amount of compromising ASes at the beginning of our dataset, \emph{without considering any updates}. On each BGP session $s_i$, we computed and maintained the routing table used to forward Tor traffic by considering all the BGP announcements and withdrawals received over $s_i$. More precisely, we kept track of the most-specific routing table entry that was used to forward traffic to any Tor guard or exit relays. We refer to those as \emph{Tor prefixes}. In this context, a routing table entry for a relay $r$ is a five-tuple $(t_i, t_f, p, e, L)$ composed of: \emph{i)} the initial time $t_i$ at which the entry started to be used by the router for forwarding traffic to $r$; \emph{ii)} the final time $t_f$ at which the entry stopped to be used by the router; \emph{iii)} the corresponding IP prefix $p$; \emph{iv)} a boolean $e$ denoting whether the $r$ is an entry or an exit relay; and \emph{v)} the list of all the ASes $L$ that will see the traffic en-route to reach $r$ (\emph{i.e.}, the \texttt{AS-PATH}).

\begin{figure}[h]
 \centering
 \includegraphics[width=0.75\columnwidth]{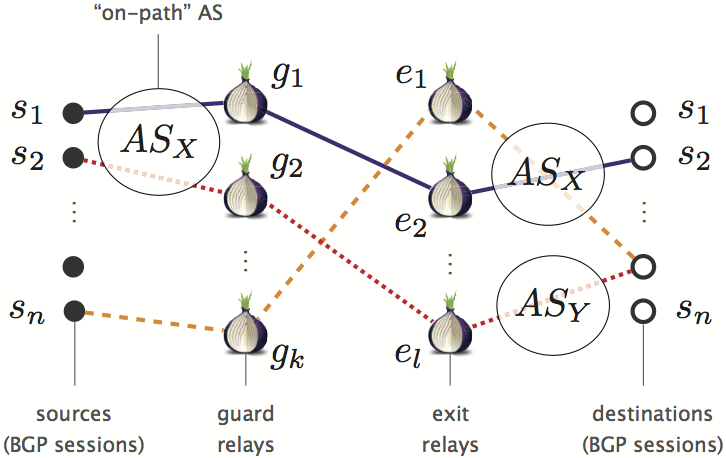}
 \caption{Control-plane evaluation setup}
 \label{fig:compromising_metric}
\end{figure}

Using the routing-table data, we accounted, for each AS $X$, the number of 
pairs (($s_i$, $g_i$), ($s_j$, $e_j$)) for which it appeared simultaneously 
in the \texttt{AS-PATH}. Here, $s_i$ (resp. $s_j$) refers to a client 
(resp. destination) session, while $g_i$ (resp. $e_j$) refers to a Tor guard (resp. exit) relay. To ensure meaningful results, we only considered cases in which $s_i$ and $s_j$ are in different ASes to ensure enough diversity in the paths seen. As illustration, in Fig.~\ref{fig:compromising_metric}, $AS_X$ is a compromising AS for the pair (($s_1$, $g_1$), ($s_2$, $e_2$)), meaning it can deanonymize any clients connected beyond $s_1$ and exchanging data with a destination connected beyond $s_2$ which uses $g_1$ (resp. $e_2$) as a guard (resp. exit) relay.

\begin{figure}[h]
 \centering
 \begin{subfigure}[t]{0.48\columnwidth}
     \raisebox{1mm}{\includegraphics[width=\columnwidth]{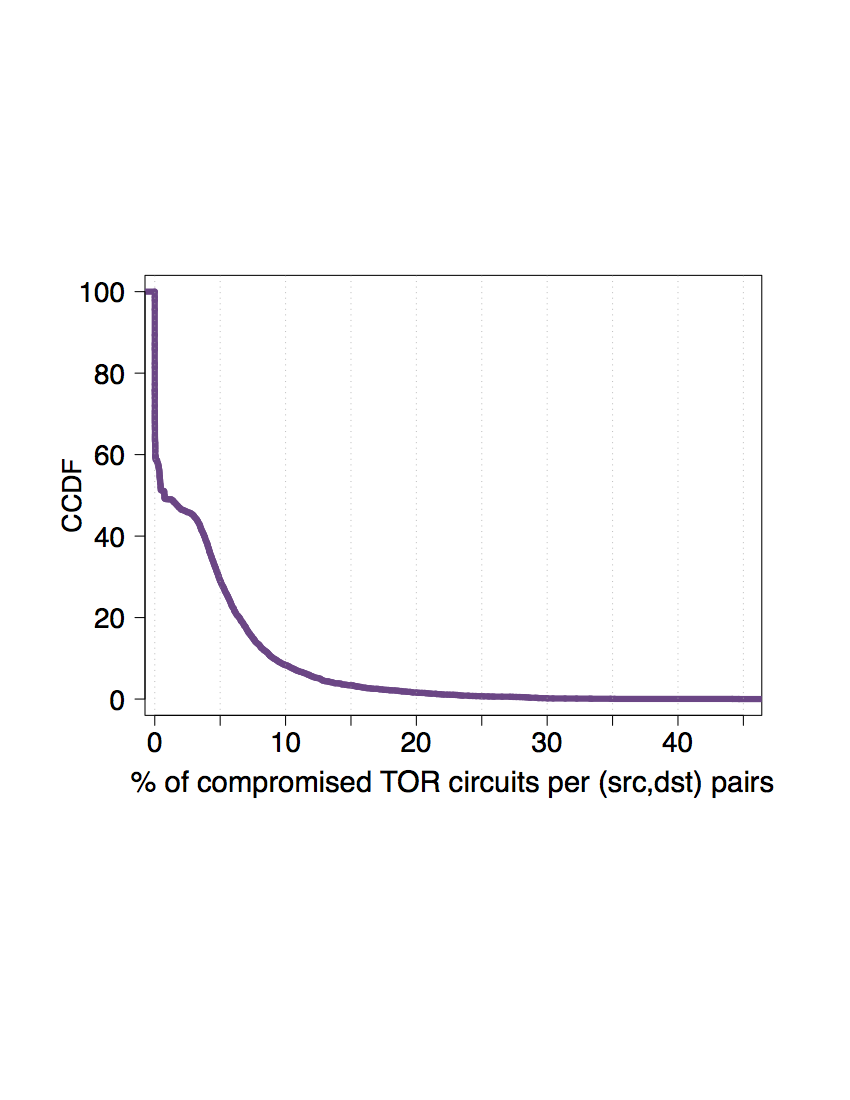}}
     \caption{Static baseline}
     \label{fig:churn_left}
 \end{subfigure}
 \hfill
 \begin{subfigure}[t]{0.50\columnwidth}
     \includegraphics[width=\columnwidth]{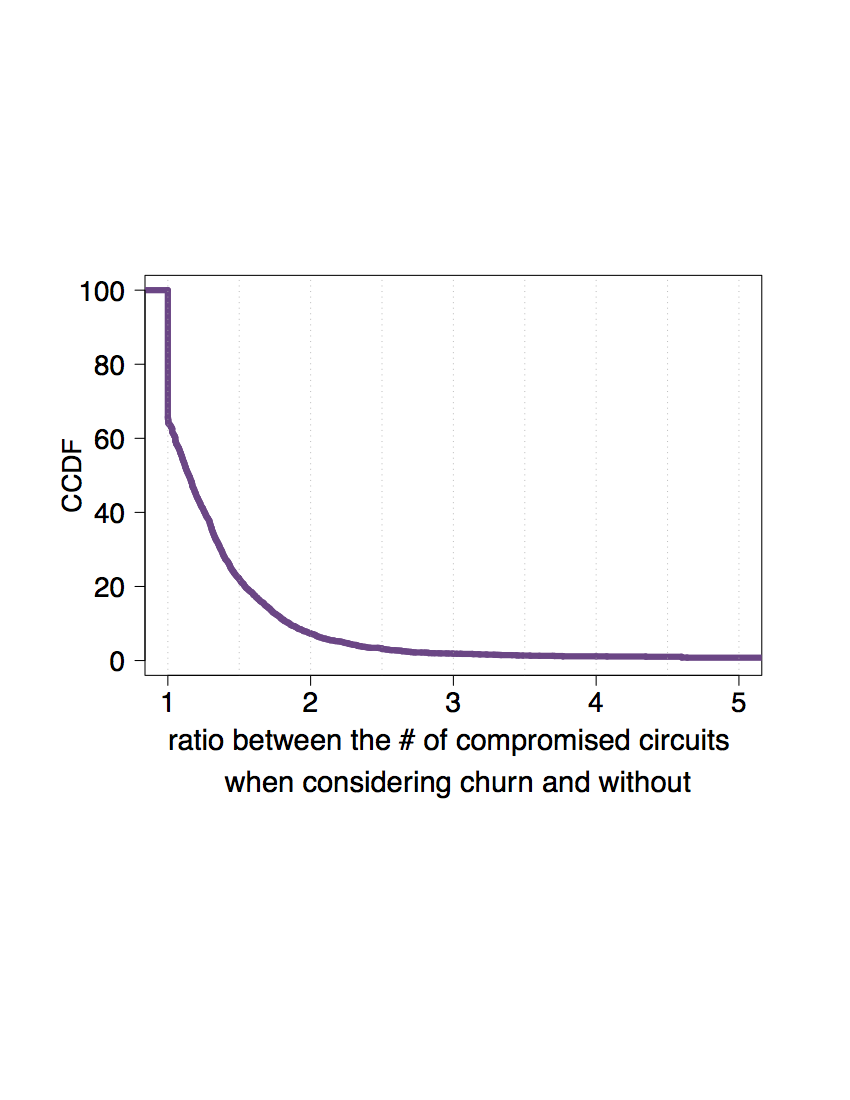}
     \caption{Churn-induced increase}
     \label{fig:churn_right}
 \end{subfigure}
 \vspace{-7px}
 \caption{Without considering churn, more than 5\% of all the possible Tor circuits are compromised by at least one AS in 20\% of the cases (left). The amount of compromised circuits increase for the majority of the (src, dest) pairs (60\%) when considering churn, by up to 50\% in 20\% of the cases (right).}
 \label{fig:churn}
\end{figure}

Fig.~\ref{fig:churn_left} depicts the percentage of compromised Tor circuits for each source and destination as a Complementary Cumulative Distribution Function (CCDF). A point $(x, y)$ on the curve means that $x$\% of all Tor circuits, \emph{i.e.} \texttt{(guard,exit)} pairs, are compromised for at least $y$\% of all the \texttt{(src,dst)} pairs. We see that, for 50\% of all the \texttt{(src,dst)} pairs, \emph{at least} 0.75\% of the Tor circuits are compromised by at least one AS. This number grows to 6\% and 13\% of the Tor circuits considering the $75^{\text{th}}$- and $95^{\text{th}}$-percentile, respectively.

\noindent\textbf{Measuring the effect of churn.} We computed the number of extra Tor circuits that got 
compromised by at least one AS over one month. To be fair, we only considered a Tor circuit as compromised 
if it crossed the same AS for at least 30 seconds as it is unlikely that a time-correlation attack can 
be performed in shorter timescale. Fig.~\ref{fig:churn_right} plots the ratio between the amount of 
compromised Tor circuits for each \texttt{(src,dst)} pair at the end of the month with respect to 
the static baseline amount. We see that churn significantly increases the probability of compromise. Indeed, the amount 
of compromised circuits increase for 60\% of the \texttt{(src,dst)}. The increase reaches 50\% (ratio of $1.5$) in 20\% of the cases.

In addition to increasing the number of compromised Tor circuits, churn also increases the number of 
compromisable \texttt{(src,dst)} pairs. Indeed, while 5593 \texttt{(src,dst)} pairs could be compromised 
without updates, that number increased to 5754 pairs when considering updates (an augmentation of nearly 3\%).

\begin{table}[h]
\centering
\scriptsize
\def\arraystretch{1.1}
\begin{tabular}{@{}llcl@{}}\toprule
Name & ASN & Tor circuits (\%) seen & Country \\
\midrule
NTT          & $2914$  & $91$ & US \\
IIJ          & $2497$  & $91$ & Japan \\
BroadbandONE & $19151$ & $91$ & US \\
Inet7        & $13030$ & $91$ & CH \\
Level3       & $3356$  & $88$ & US \\
Tinet        & $3257$  & $86$ & DE \\
Cogent       & $174$   & $63$ & US \\
Level3/GBLX  & $3549$  & $58$ & US \\
TATA AMERICA & $6453$  & $53$ & US \\
TeliaSonera  & $1299$  & $50$ & SWE \\
\bottomrule
\end{tabular}
\caption{A few well-established ASes simultaneously see some traffic for up to 90\% of all (entry, exit) relays pairs.}
\label{tab:ases_tor_circuits}
\end{table}

\noindent\textbf{Few powerful ASes see some traffic for a large majority of the Tor circuits.} Due to their 
central position in the Internet, a few ASes naturally tend to see a lot of Tor traffic crossing them. 
To account for this effect, we compute how many Tor circuits crossed each AS from at least 
one \texttt{(src,dst)} pair. The top 10 ASes in terms of compromised circuits are listed in 
Table~\ref{tab:ases_tor_circuits}. Large networks such as NTT or Level3 are able to see Tor traffic for 
up to 90\% of Tor circuits.

\subsection{Data-plane Evaluation}
\label{ssec:dp_churn_eval}

Next, we aim to quantify the impact of churn using data-plane information collected via \texttt{traceroute}. 

\noindent\textbf{Datasets} We ran \texttt{traceroute} between 70 RIPE Atlas probes~\cite{ripe_atlas} to measure the actual forwarding path taken by packets entering and exiting the Tor network. We selected one probe in 70 different ASes, split in the following four sets:
\begin{itemize}[leftmargin=*]
\setlength{\itemsep}{0pt}
\item $S_1:$ 10 ASes that contain the most Tor clients~\cite{juen};
\item $S_2:$ 25 ASes that cumulatively contained {\raise.17ex\hbox{$\scriptstyle\sim$}}50\% of all guard relay bandwidth;
\item $S_3:$ 25 ASes that cumulatively contained {\raise.17ex\hbox{$\scriptstyle\sim$}}50\% of all exit relay bandwidth;
\item $S_4:$ 10 ASes that contain the most Tor destinations~\cite{juen}.
\end{itemize}
We then ran daily traceroutes over a 3 weeks period between all probes in $S_1$ towards all probes in $S_2$ (and vice-versa), measuring the forwarding paths $P_1$ between Tor clients and guard relays, and the paths $P_2$ between guard relays and Tor clients. Similarly, we measured the forwarding paths $P_3$ between exit relays and Tor destinations, and the paths $P_4$ between Tor destinations and exit relays. Overall, we measured $10 \times 25 \times 25 \times 10 = 62500$ possible Tor circuits.

\begin{figure}[h]
\centering
\includegraphics[width=.75\linewidth]{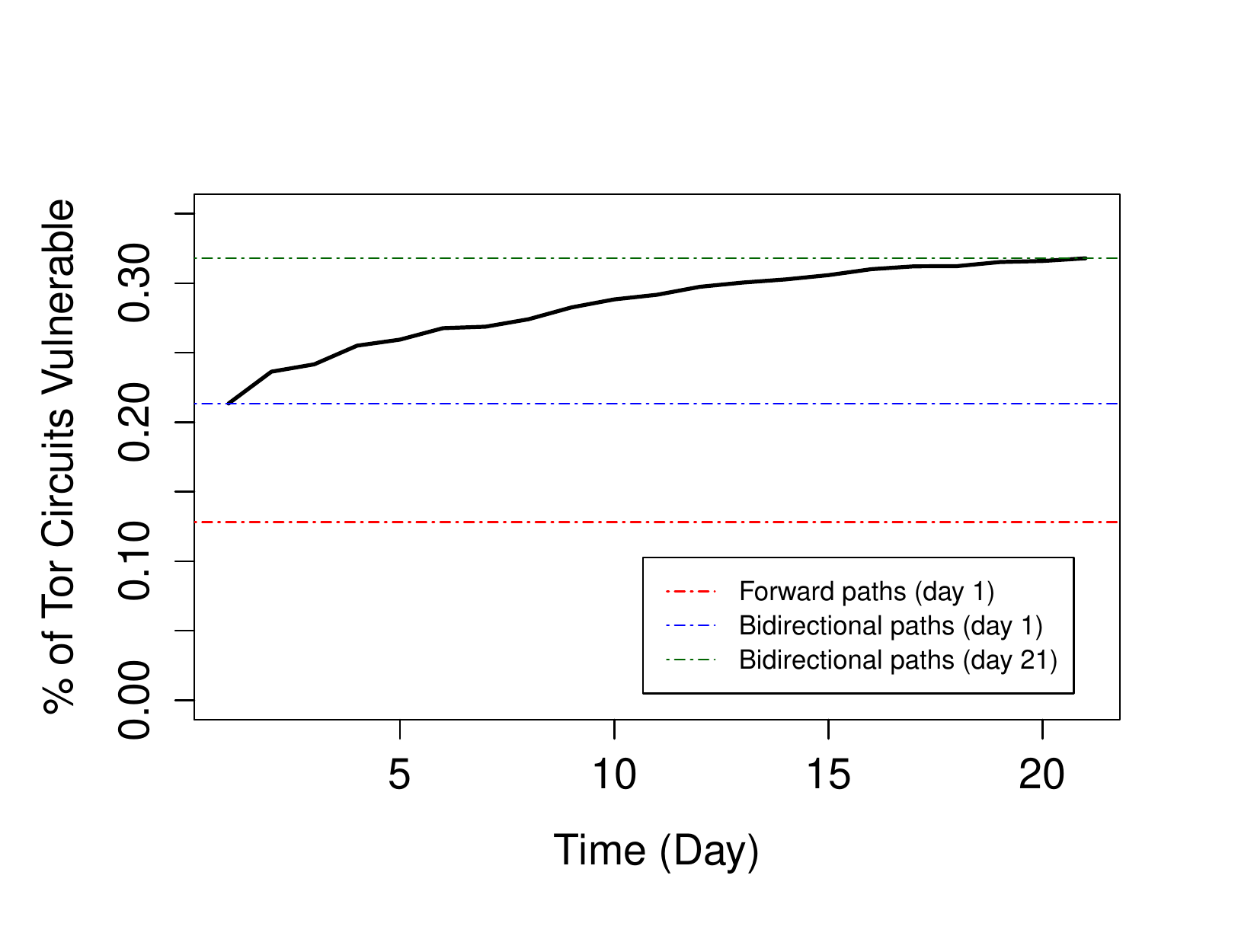}
\caption{Percentage of Tor circuits vulnerable to an AS level adversary}
\label{fig:vulnerability_time}
\end{figure}

\noindent\textbf{Churn nearly tripled the amount of vulnerable Tor circuits.} If we use conventional methodology 
and only look for common ASes between $P_1$ and $P_3$, we found 12.8\% of Tor circuits to be vulnerable on the first 
day of the experiment (red line in Fig.~\ref{fig:vulnerability_time}). In comparison, if we also consider 
asymmetric paths (\emph{i.e.}, also look for common ASes between $P_1$ and $P_4$, $P_2$ and $P_3$, and $P_2$ and $P_4$),
 the percentage of vulnerable Tor circuits nearly doubled to 21.3\% on the first day (blue line in Fig.~\ref{fig:vulnerability_time}), 
and nearly tripled to 31.8\% at the end of the three week period (green line in Fig.~\ref{fig:vulnerability_time}).

\section{BGP Attacks: Hijack and Interception}\label{sec:hijacks_and_interception}

In this section, we study and evaluate the feasibility of BGP hijack and interception attacks 
on the Tor network. First, we show that Tor relays tend to be concentrated within few ASes and IP prefixes---making those highly attractive targets for hijack and interception attacks (\S\ref{ssec:high_concentration}). 
Second, we show that, in several real-world BGP hijack attacks, Tor relays were among the target prefixes (\S\ref{sec:knownattacks}).
Third, we perform a real-world BGP interception attack against a live Tor guard relay, \emph{with success}, 
to demonstrate the ability to accurately deanonymize Tor clients (\S\ref{ssec:interception_attack}). 

\subsection{Tor relays concentration}
\label{ssec:high_concentration}

The amount of Tor traffic attracted by a hijack or an interception attack depends on the number of relays that lie within the corresponding prefix. As such, prefixes and ASes that host many relays of high bandwidth are an interesting targets for attackers. To evaluate how vulnerable the Tor network was to hijack and interceptions attacks, we computed the number of ASes present in each AS and in each BGP prefix.  Surprisingly, close to 30\% of all relays are hosted in only 6 ASes and 70 prefixes. Together, these relays represent almost 40\% of the bandwidth in the entire Tor network (see Table~\ref{tab:ases_concentration}). As such, these few prefixes constitute \emph{extremely} attractive targets.
 
\begin{table}
\centering
\footnotesize
\begin{tabular}[b]{@{}lccclll@{}}\cmidrule[0.08em]{2-6}
& \% relays & \% bw & \# pfx & Name & ASN \\
\cmidrule[0.05em]{2-6}
        & 10.5 & 23 & 11.80   & OVH            & 16276  \\
        & 6.30 & 13 & 6.68    & Hetzner        & 24940  \\
        & 4.78 & 7  & 10.52   & Online.net     & 12876  \\
        & 3.04 & 4  & 2.58    & Wedos          & 197019 \\
        & 2.04 & 14 & 4.27    & Leaseweb       & 16265  \\
        & 1.69 & 9  & 3.86    & PlusServer     & 8972   \\
\cmidrule[0.05em]{2-6}
Total   & 28.35 & 70 & 39.71 & & \\
\cmidrule[0.08em]{2-6}
\end{tabular}
\caption{6 ASes and 70 prefixes host {\raise.17ex\hbox{$\scriptstyle\sim$}}30\% of all Tor guard and exit relays as well as {\raise.17ex\hbox{$\scriptstyle\sim$}}40\% of the entire Tor network bandwidth. As such, these constitute extremely attractive targets for hijacks and interceptions attacks.}
\label{tab:ases_concentration}
\end{table}

\subsection{Known Prefix Hijacking Attacks}
\label{sec:knownattacks}
While there have been numerous well-documented BGP prefix hijacks and interceptions, it was unknown whether Tor traffic was intercepted or not. To this extent, we studied occurrences of well-known prefix hijacks and looked for leaked prefixes covering at least a Tor relay. To do so, we gathered BGP updates from Routeviews~\cite{routeviews} around the time of each attack and filtered out the ones related to Tor prefixes. Overall, we found that three well-known hijacks affected Tor relays: two separate incidents involving one of Indonesia's largest telecommunication networks, Indosat, as well as one malicious hijack attack whose goal was to steal Bitcoins.  

\begin{table}[ht]
\centering
\def\arraystretch{1.1}
  \begin{tabular}{@{}lllllll@{}}\toprule
  \multirow{2}{*}{Event} & \# hijacked & \# hijacked  & \# hijacked  \\
  & relays & guards & exits \\
  \midrule
  Indosat 2011 & \multicolumn{1}{c}{5 (0.24\%)}  & \multicolumn{1}{c}{1 (0.15\%)} & \multicolumn{1}{c}{4 (0.44\%)}\\
  Indosat 2014 & \multicolumn{1}{c}{44 (0.80\%)} & \multicolumn{1}{c}{38 (1.80\%)} & \multicolumn{1}{c}{17 (1.65\%)}\\
  \bottomrule
  \end{tabular}
\caption{Summary statistics for known Indosat prefix hijacking events.}
\label{tab:indosat}
\end{table}

\noindent\textbf{Indosat 2011.}  On January 14th, 2011, Indosat (AS4761) originated 2,800 new prefixes, which covered 824 
different ASes~\cite{indosat}. 7 of these prefixes affected the Tor network by covering 5 of the Tor relays.  
As discussed in Section~\ref{sec:raptor}, Indosat \emph{could} potentially have learned information about the 
client IP addresses associated with each of the guard relays (reduced anonymity set). 

\noindent\textbf {Indosat 2014.} On April 3, 2014, Indosat originated 417,038 new prefixes; it usually originates 300 
prefixes~\cite{indosat2014}.  This compromised 44 Tor relays, 38 of which were 
guard relays and 17 of which were exit relays (11 hijacked relays were both guards and exits).  
Table~\ref{tab:indosat} shows the summary statistics of both Indosat hijacking incidents. 

\noindent\textbf{Canadian Bitcoin 2014.}  From February 2014 to May 2014, an attacker compromised 51 networks at 19 different 
ISPs, and resulted in the theft of approximately \$83,000 in Bitcoin~\cite{bitcoin}.  We found that 198.245.63.0/24 and 
162.243.142.0/24 were hijacked, and contained a Tor relay, 198.245.63.228.  AS16276 (OVH) owns 198.245.63.0/24, but this 
prefix was hijacked by AS21548 (MTO Telecom).  The Tor relay that consequently was hijacked, 198.245.63.228, was a guard 
relay located in Montreal, Quebec.  

While we do not make any claims about the intent of the above hijacking ASes, our analysis 
shows the existential threat of real-world routing attacks on the Tor network. Furthermore, 
the fact that the Tor and the research community missed noticing the presence of Tor relays among 
the hijacked prefixes is surprising.

\subsection{BGP Prefix Interception Attack Experiment}
\label{ssec:interception_attack}

\noindent\textbf{Methodology and setup.} We now demonstrate the feasibility of the interception attack by performing one, with success, on the live Tor network. For that, we set up a machine to run as a Tor guard relay and made it reachable to the Internet by announcing a \texttt{/23} prefix in BGP using Transit Portal (TP)~\cite{Schlinker:2014:PU:2670518.2673887}. TP enables virtual ASes to establish full BGP connectivity with the rest of the Internet by proxying their announcements via dozens of worldwide deployments. Next, we configure the $50$ Tor clients in PlanetLab to use our Tor guard relay as the entry relay to reach $50$ web servers, also hosted in PlanetLab.

\begin{figure}
\centering
\includegraphics[trim=30 100 30 10, clip, width=1\linewidth]{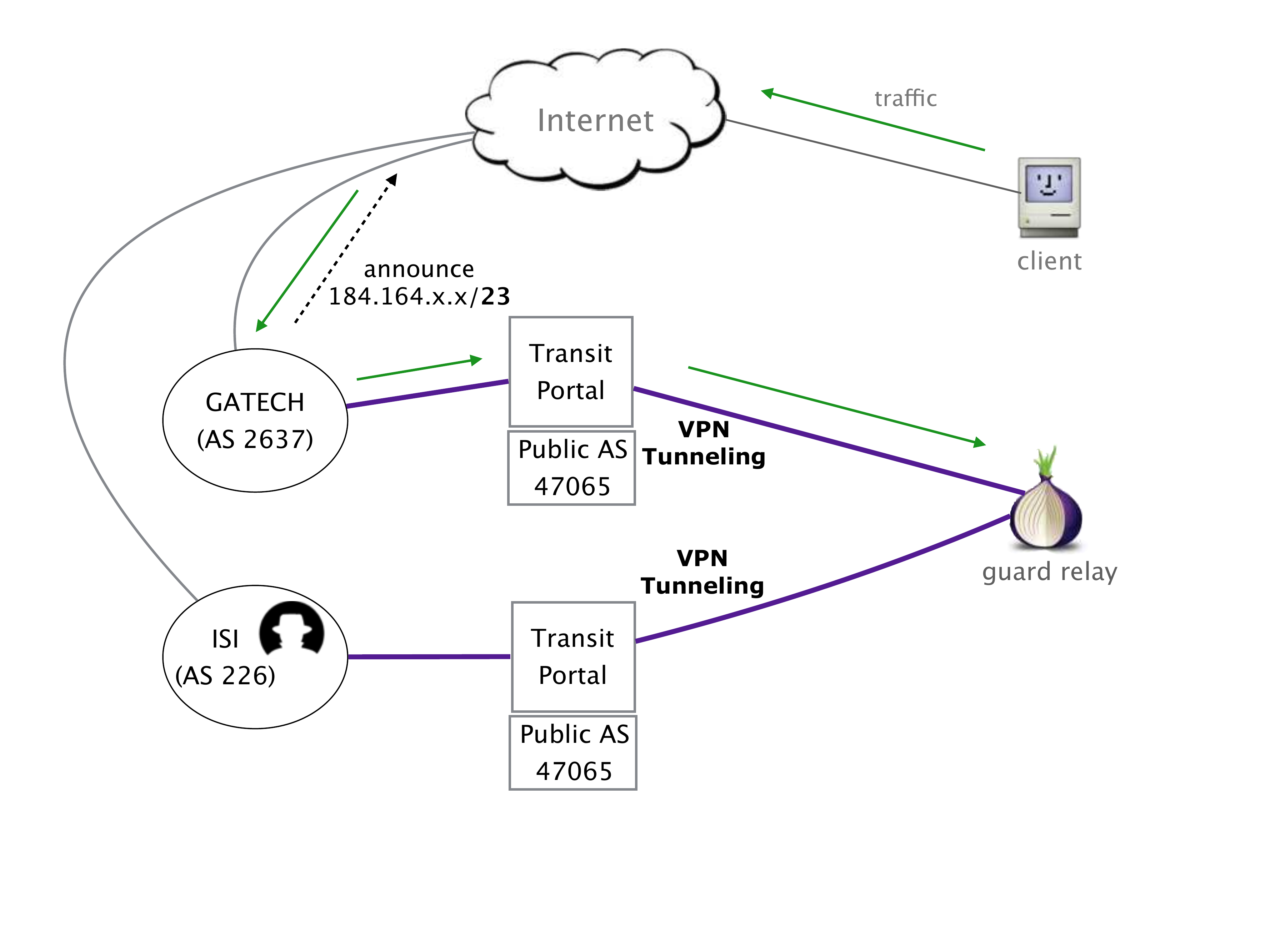}
\caption{Transit Portal setup}
\label{fig:intercept_setup}
\end{figure}

In order to perform the BGP prefix interception attack, we used two TP deployments (\texttt{GATECH} and \texttt{ISI}), located in different ASes. GATECH TP served the purpose of the ``good'' AS through which Tor traffic is normally routed, while ISI TP served as the ``malicious'' AS which performed the interception attack. We connected the two TPs to our Tor relay machine via VPN tunnels. First, in order for our Tor guard relay (running on \texttt{184.164.x.x}) to be reachable, we advertised \texttt{184.164.x.x/23} via the GATECH TP, so that traffic destined for IP addresses within that range will be routed, first to the GATECH TP, and then sent to our machine via the corresponding tunnel. We illustrate our setup in Fig.~\ref{fig:intercept_setup}.

\begin{figure}
\centering
\includegraphics[trim=30 100 30 10, clip, width=1\linewidth]{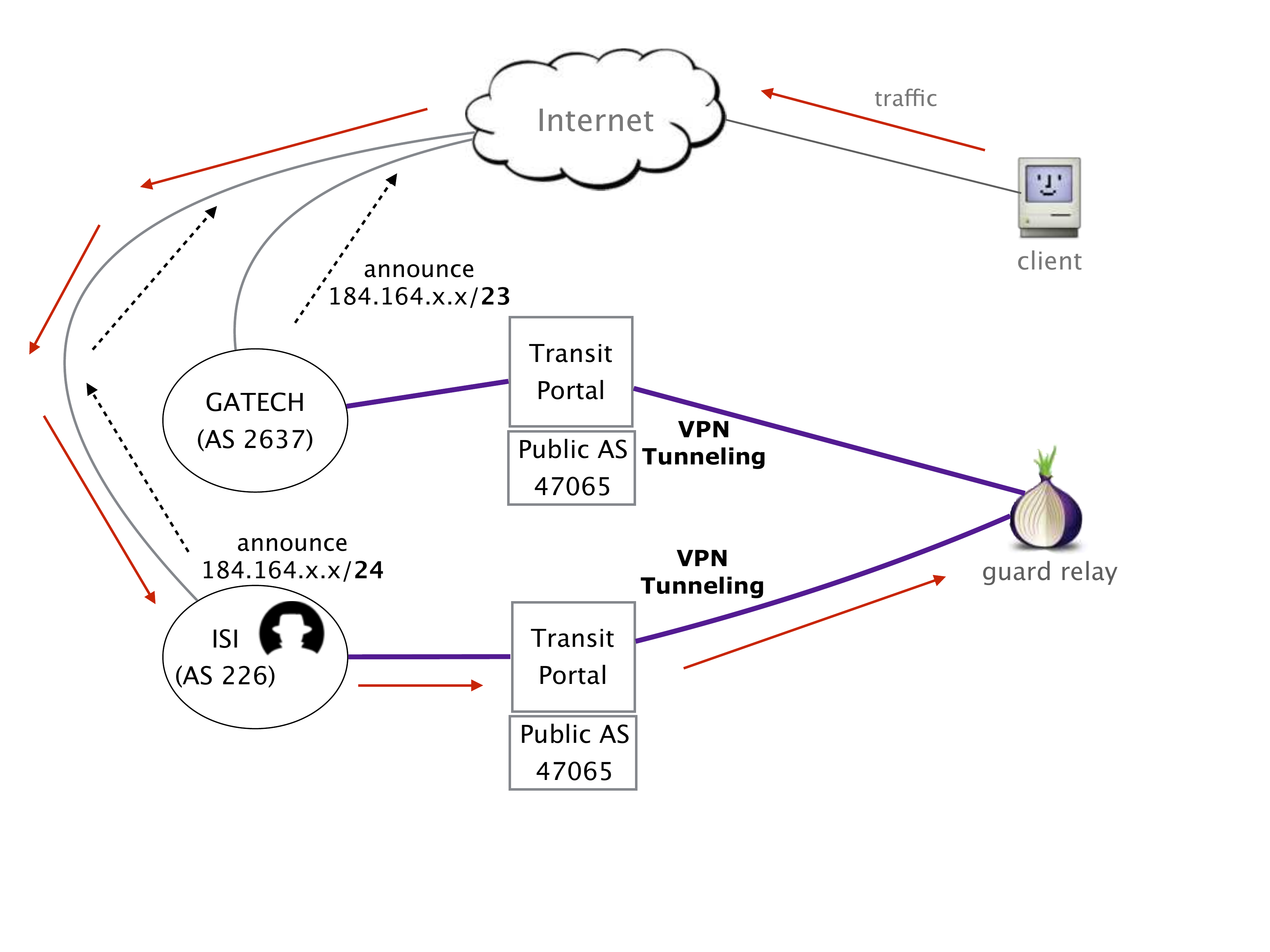}
\caption{ISI performing interception attack}
\label{fig:intercept_attack}
\end{figure}

Next, We advertise BGP prefix \texttt{184.164.x.x/24} via the ISI TP, which constitutes a more-specific prefix attack against the original announcement announced by the GATECH TP. Thus, after the new BGP prefix announcement gets propagated through the internet, Tor client traffic that is destined for our guard will be sent to ISI instead. Since we configured the ISI TP to forward traffic to our guard machine, the Tor relay can still receive the traffic and keep the Tor connection alive after the attack. We illustrate the interception attack model in Fig.~\ref{fig:intercept_attack}.

Our setup constitutes a BGP interception attack. Initially, traffic is routed via GATECH and arrives at our Tor relay machine via GATECH tunnel. After the attack happens, traffic drains from GATECH tunnel and gets routed via ISI, and thus comes to our Tor relay machine via ISI tunnel instead. Since the traffic still arrives at the relay machine, it is an interception attack and the connection does not get interrupted. We use \texttt{tcpdump} on our relay machine, listening to ISI tunnel, to capture client TCP acknowledgment traffic coming from that tunnel, which is exactly the data that an adversary would get from launching such an interception attack.

In the experiment, we first launch simultaneous HTTP requests using \texttt{wget} at the $50$ Tor clients for the $100$MB file at the $50$ web servers. Then, $20$ seconds after launching the \texttt{wget} requests, we start announcing the more-specific prefix via ISI. We use \texttt{tcpdump} listening to ISI tunnel to capture TCP acknowledgment traffic sent from the Tor clients during the interception attack. We also use \texttt{tcpdump} to capture traffic at the web servers during the whole process. Finally, $300$ seconds after launching the attack, we send a withdrawal message via the ISI TP, so the traffic will be routed via GATECH again as normal.

\begin{figure}
\begin{subfigure}{.25\textwidth}
  \centering
  \includegraphics[width=.9\linewidth]{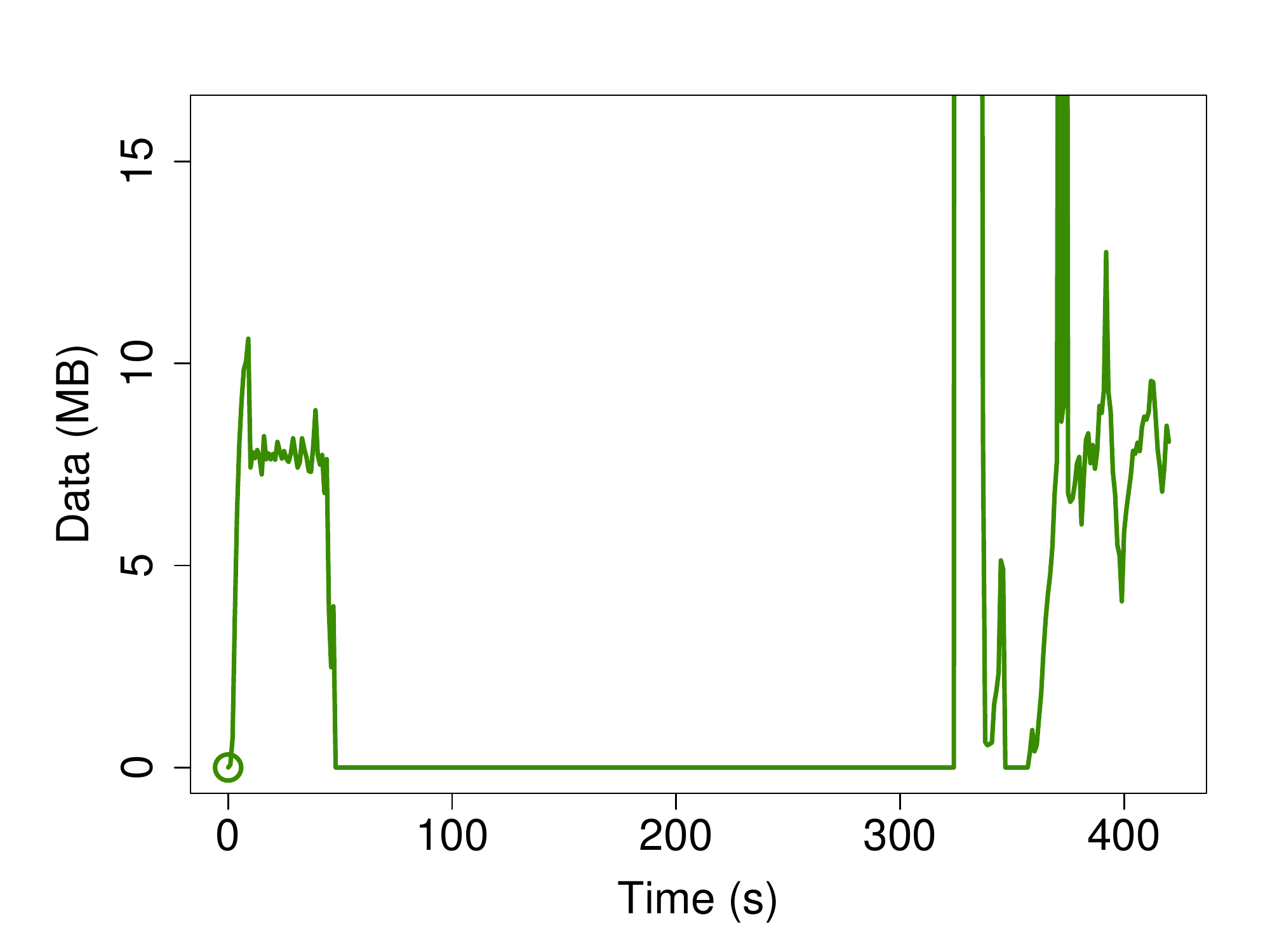}
  \caption{Traffic Flow via GATECH}
\end{subfigure}%
\begin{subfigure}{.25\textwidth}
  \centering
  \includegraphics[width=.9\linewidth]{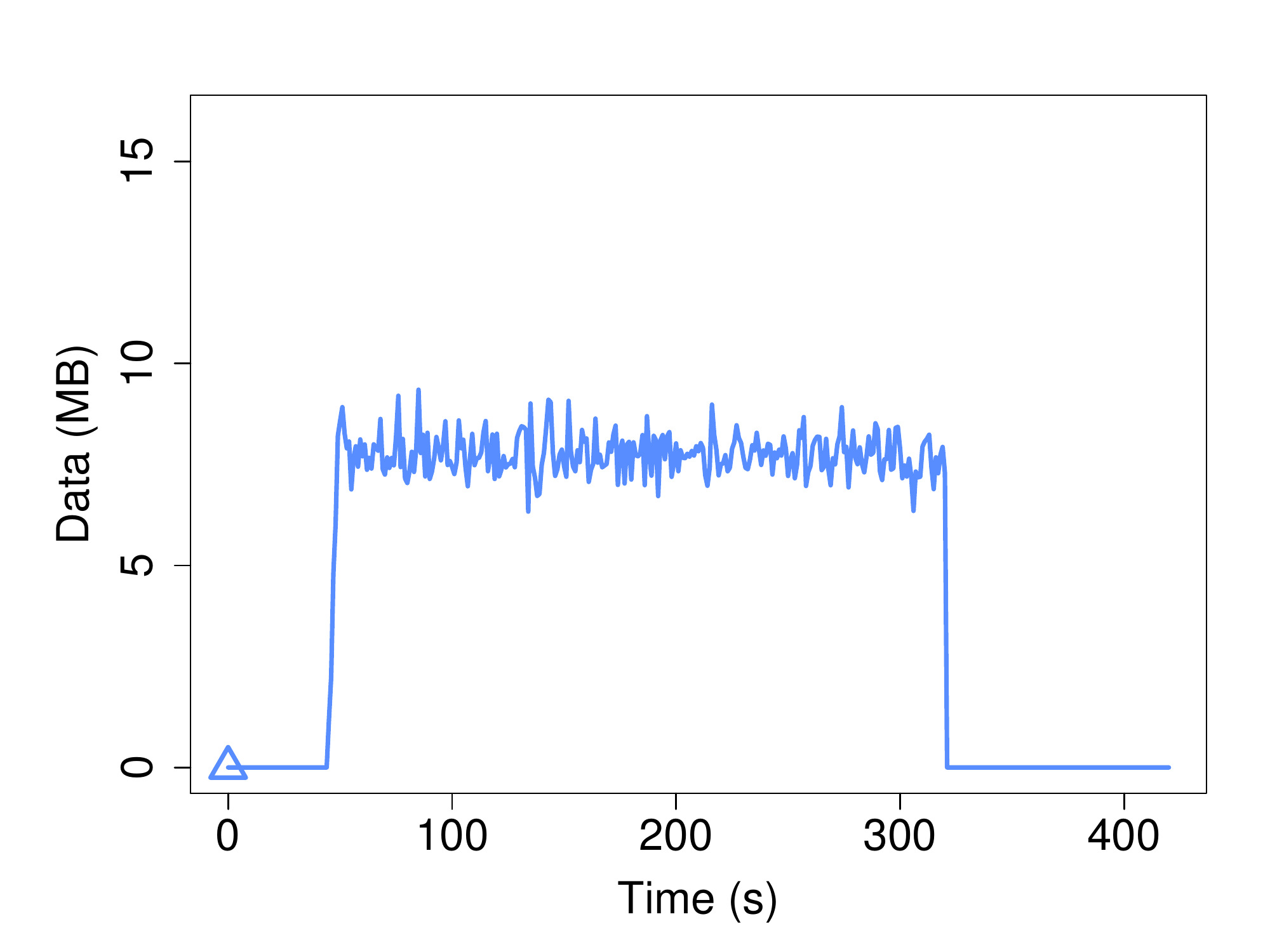}
  \caption{Traffic Flow via ISI}
\end{subfigure}
\caption{Traffic Flow During the Experiment}
\label{fig:intercept_relay}
\end{figure}

\noindent\textbf{Our interception attack successfully deanonymized Tor sources with a $90\%$ accuracy rate.} In Fig.~\ref{fig:intercept_relay}, we plot the Tor traffic flow captured on our relay machine from both GATECH tunnel and ISI tunnel. We can see that all traffic is routed via GATECH at the beginning. At $t=20s$, ISI starts advertising a more specific $/24$ prefix, which takes approximately $35$ seconds for it to be propagated through the internet and drain the traffic from GATECH. At $t=55s$, traffic starts showing up via ISI, and GATECH does not receive traffic any more. Then, at $t=300s$, ISI withdraws the IP prefix announcement, which takes approximately $22$ seconds for the traffic to appear back on GATECH again. During this interception process, the connection stays alive.

The captured data from ISI tunnel is client TCP acknowledgment traffic. Thus, we will employ our Asymmetric Traffic Analysis approach, described in Section~\ref{sec:asymmetry}, to do the correlation analysis to deanonymize users' identity. We achieve $90\%$ accuracy rate (see Table~\ref{tbl:planetlab-intercept}).

\begin{table}[ht]
\scriptsize
\centering
\def\arraystretch{1.1}
  \begin{tabular}{@{}lccc@{}}\toprule
  & Accuracy & False & False \\
  & Rate & Negative & Positive \\
  \midrule
  Client ACK/Server ACK    & 90\% & 8\% & 2\% \\
  \bottomrule
  \end{tabular}
\caption{Asymmetric Traffic Analysis accuracy rate}
\label{tbl:planetlab-intercept}
\end{table}

\begin{figure}
\begin{subfigure}{.25\textwidth}
  \centering
  \includegraphics[width=.9\linewidth]{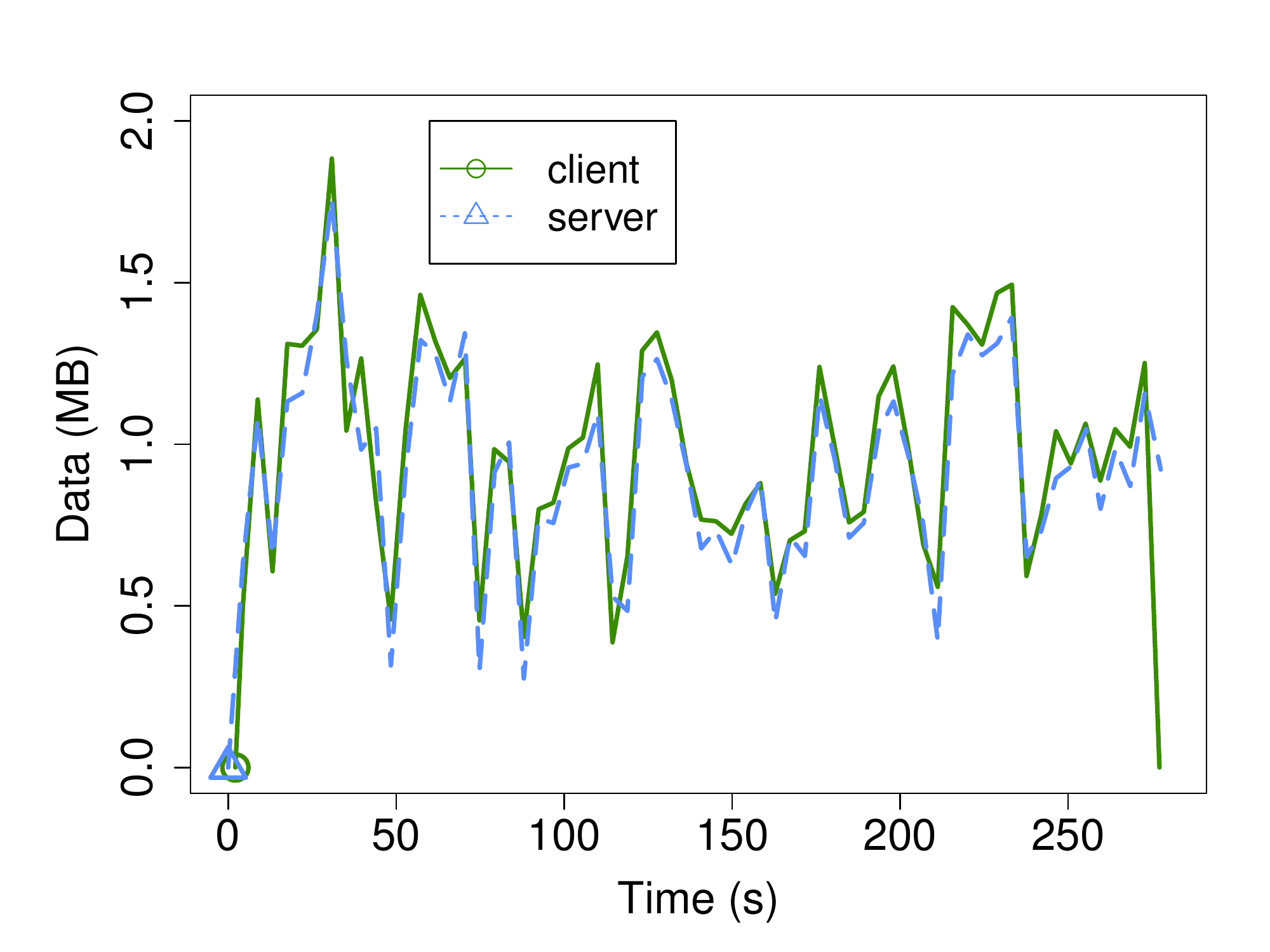}
  \caption{Client \& Correlated Server}
\end{subfigure}%
\begin{subfigure}{.25\textwidth}
  \centering
  \includegraphics[width=.9\linewidth]{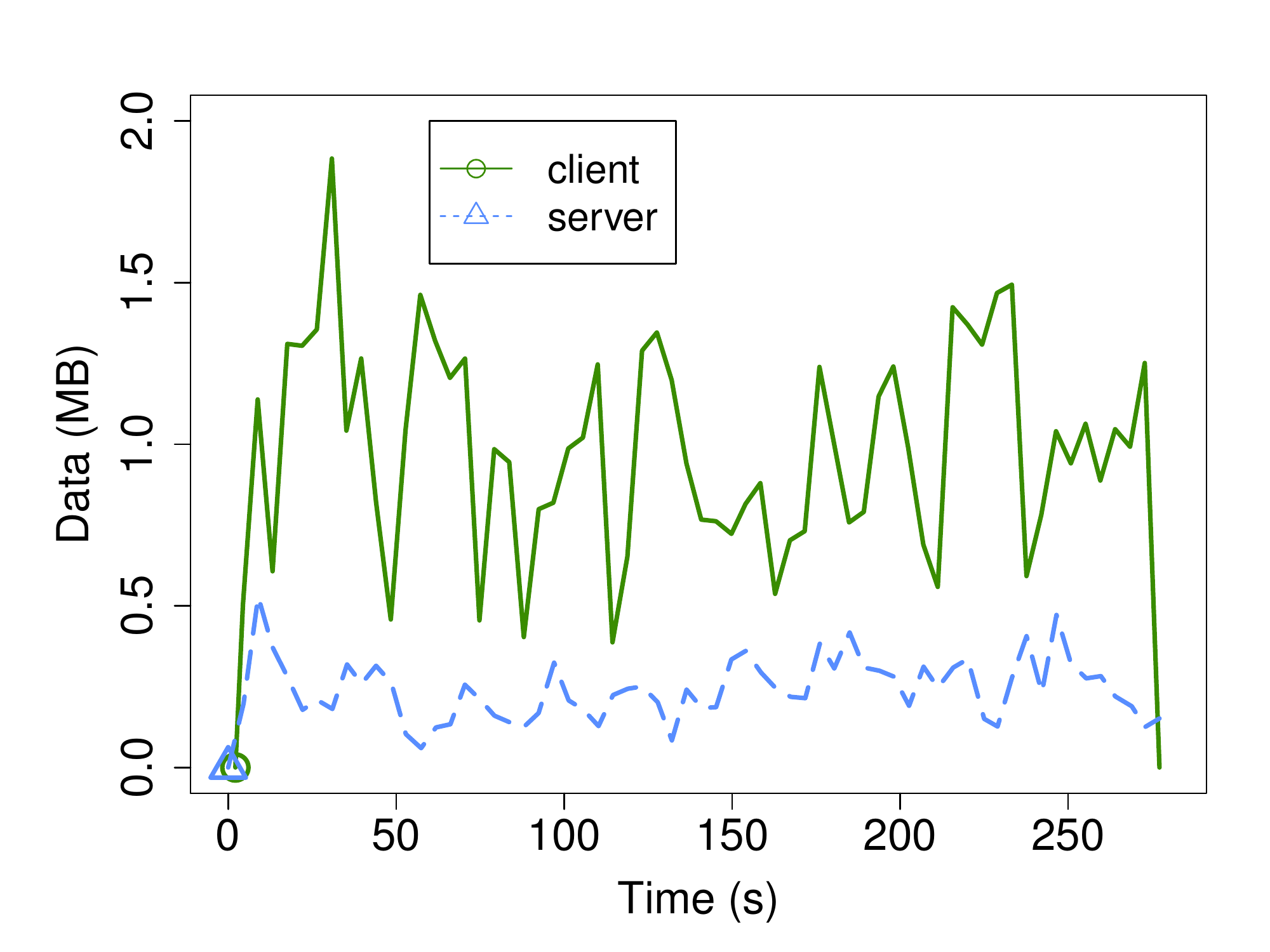}
  \caption{Client \& Uncorrelated Server}
\end{subfigure}
\caption{Client ACK versus Server ACK analysis}
\label{fig:intercept_corr}
\end{figure}

Fig.~\ref{fig:intercept_corr} shows an example of a client with its correlated server and an uncorrelated server, respectively. Note that the time shown on the graph has been adjusted according to the time that traffic starts showing via ISI.

The detection accuracy rate in the interception attack case decreases from the average $95\%$ in static asymmetric traffic analysis to $90\%$. One main reason is that we configure all $50$ Tor clients to connect to the same Tor guard relay, which leads to significantly higher probability that many of them will share the same Tor exit relay (especially those clients which are in the same AS) as well, and as a result, their bandwidths are highly likely to be similar. And also, all the clients start requesting files from the web servers at the same time, so the bandwidth they could achieve will be limited by the guard and the exit relay, which leads to similar bandwidths due to the guard/exit bottleneck. However, this scenario is an extreme and very unlikely case in real Tor connections. With fewer clients connecting to the same Tor guard relay at the same time, the accuracy of the asymmetric traffic analysis should be higher.

\begin{figure}
 \centering
 \includegraphics[width=0.75\columnwidth]{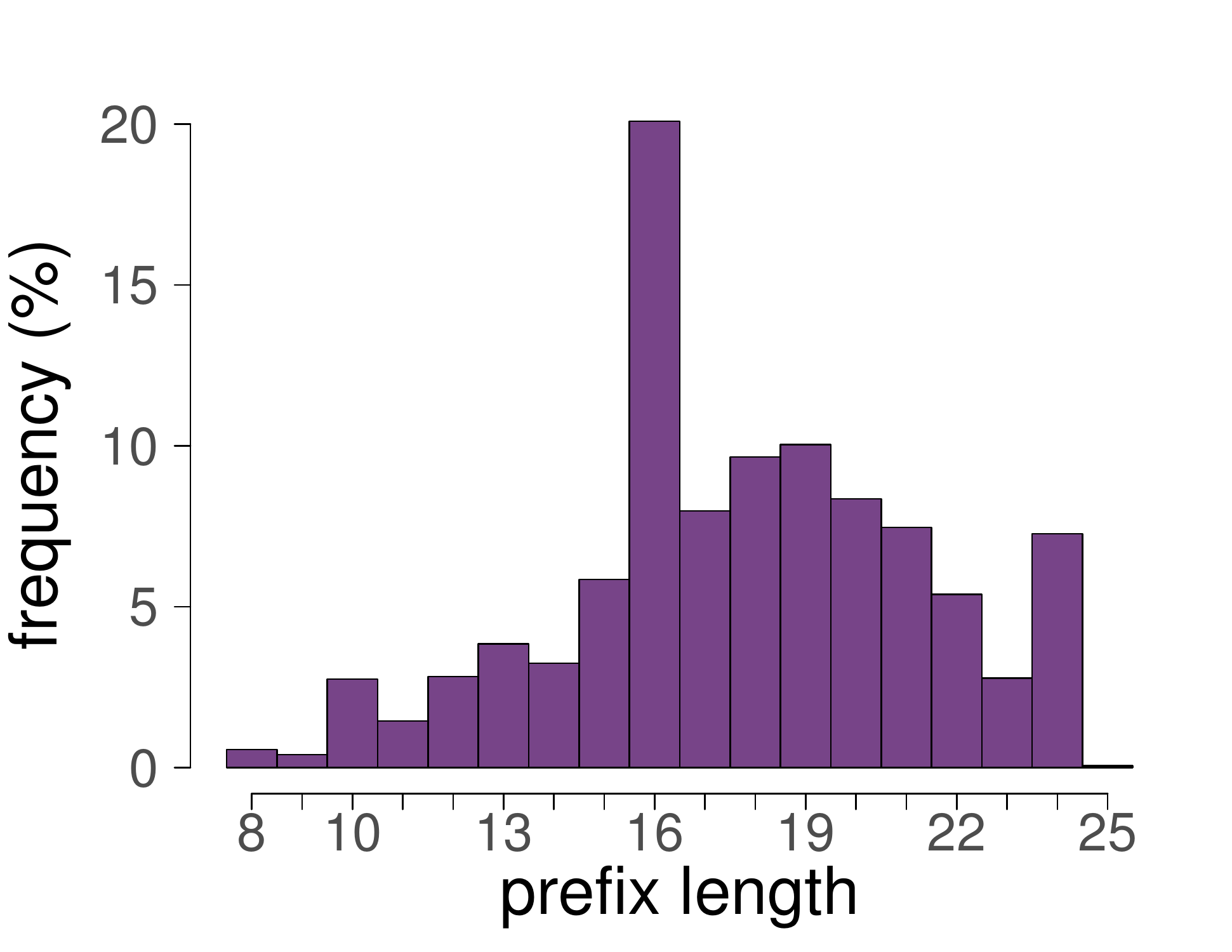}
 \caption{\textgreater 90\% of BGP prefixes hosting relays are shorter than \texttt{/24}, making them vulnerable to our attack.}
 \label{fig:prefix_length}
 \vspace{-7px}
\end{figure}

\noindent\textbf{The vast majority of Tor relays are vulnerable to our attacks.} Technically, only prefixes shorter than $/24$ can be hijacked globally with a more-specific prefix attack as longer prefixes tend to be filtered by default by many ISPs. To make sure of the feasibility of our attack, we computed the prefix length distribution of Tor prefixes (see Fig.\ref{fig:prefix_length}). We can see that \emph{more than $90\%$} of BGP prefixes hosting relays have prefix length shorter than $/24$, making them directly vulnerable to a more-specific prefix attack such as ours.

\section{Countermeasures Sketch}
\label{sec:countermeasures}

In this section, we first describe a taxonomy of countermeasures against Raptor attacks. Second, we describe a general approach for 
AS-aware anonymous communication in which Tor clients are aware of the dynamics of Internet routing. 
Finally, we describe exploratory approaches for detecting and preventing BGP hijack and interception attacks against Tor.

\subsection{Countermeasure Taxonomy}

\begin{figure}[h]
\centering
\includegraphics[width=.9\linewidth]{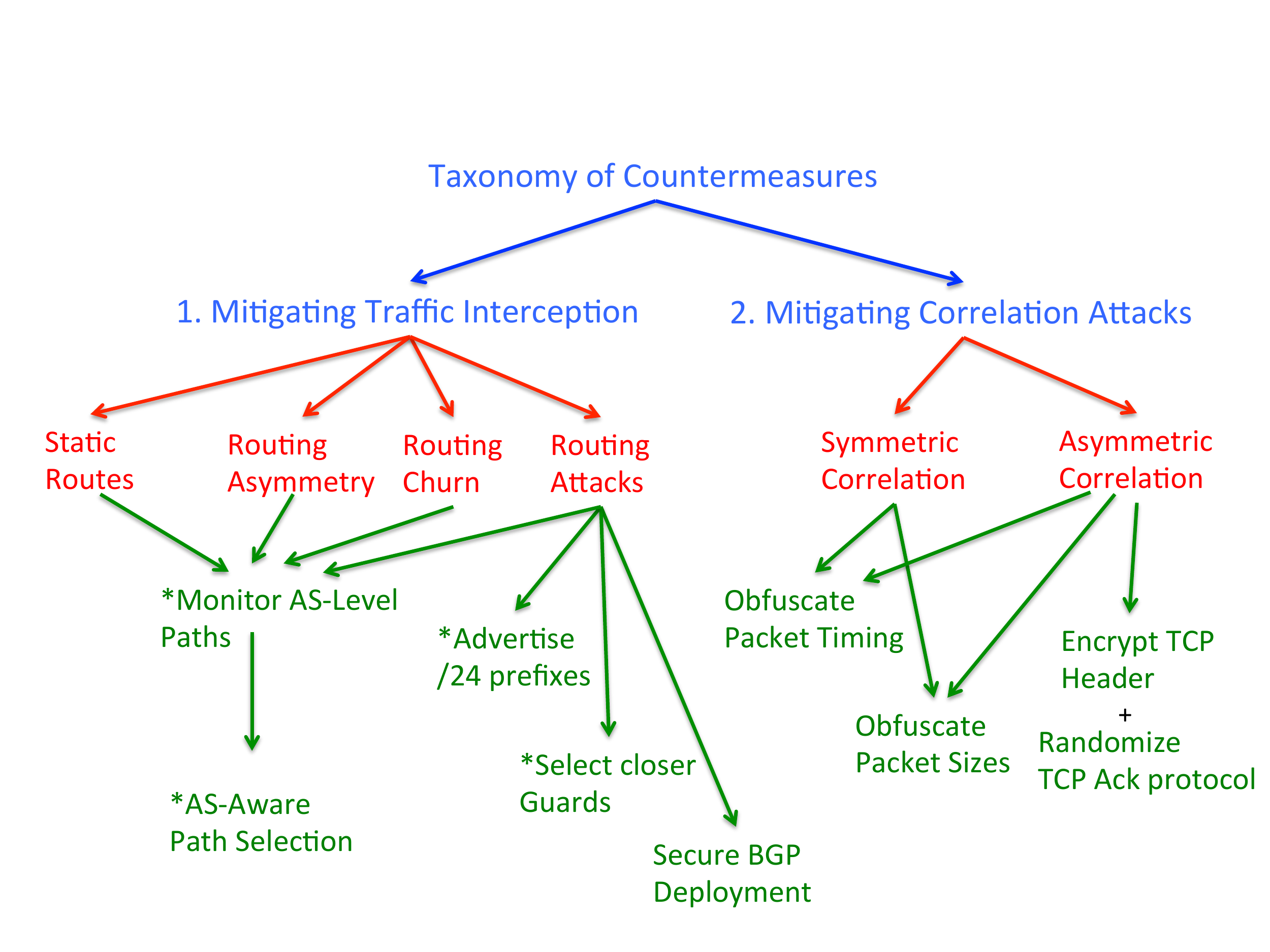}
\caption{Taxonomy of Countermeasures}
\label{fig:countermeasures}
\end{figure}

There are two main categories of countermeasures: (a) approaches that reduce the chance of an AS-level adversary 
observing both ends of the anonymous communication, and (b) approaches that aim to mitigate correlation attacks 
even when an adversary observes both ends of the anonymous communication. Figure~\ref{fig:countermeasures} 
illustrates the design space of potential countermeasures against Raptor attacks. 
In this work, we advocate the former line of defense -- namely, 
to monitor both routing control-plane and data-plane, and to strategically select Tor relays that 
minimize the chance of compromise (\S\ref{sec:as-path-selection}). We also advocate defenses that aim to detect and prevent 
routing attacks (\S\ref{sec:counter-routing-attacks}). We do not focus on the class of approaches that aim 
to mitigate correlation analysis by obfuscating packet sizes and timings, as they are generally 
considered too costly to deploy (Appendix \ref{appendix:obfuscation}).

\subsection{AS-Aware Path Selection} 
\label{sec:as-path-selection}

To minimize opportunities for AS-level traffic analysis, 
the Tor network can monitor the path dynamics between the clients 
and the guard relays, and between the exit relays and the destinations. 
Information about path dynamics can be obtained using data-plane (\emph{e.g.}, \texttt{traceroute}) 
or control-plane (\emph{e.g.}, BGP feed) tools. 
For instance, each relay could publish the list of any ASes it used to 
reach each destination prefix in the last month.
This information can be distributed 
to all Tor clients as part of the Tor network consensus data.  Tor
clients can use this data in relay selection, perhaps in
combination with their own traceroute measurements of
the forward path to each guard relay.
For example, Tor clients should select relays 
such that the same AS does not appear in both the first and 
the last segments, after \emph{taking path dynamics into account}.

\subsection{Mitigating Routing Attacks in Tor}
\label{sec:counter-routing-attacks}
Next, we consider two approaches for mitigating Raptor's 
routing attacks: detection and prevention.
 
\subsubsection{Monitoring Framework for Detection Routing Attacks}

We propose that the Tor network monitor the routing control-plane
 and data-plane for robust detection of routing attacks. 
Detecting routing attacks serves two purposes: (1) First, 
this serves to raise awareness about the problem and hold 
attackers accountable. (2) Second, Tor directory authorities 
can notify clients. Such notifications allow the end-user 
to respond by either suspending its use of Tor (since 
most hijacks and interceptions are short lived), or 
choose another Tor relay (see Appendix \ref{appendix:monitoring-challenges} 
for challenges and caveats for this approach). Next, we 
discuss two proof-of-concept monitoring frameworks, based on 
BGP data and traceroute data respectively.
 
\noindent\textbf{BGP Monitoring Framework.}
Our BGP monitoring framework gathers BGP data from the Routeviews project. The framework 
filters BGP updates to consider data about prefixes that involve a Tor relay. Building 
upon prior work in routing attack detection~\cite{ballani:sigcomm07}, we implement the 
following heuristics. (1) \emph{Frequency heuristic:}  routing attacks can be 
characterized by an AS announcing a path once (or extremely rarely) to a prefix 
that it does not own.  The frequency heuristic detects attacks that exhibit this behavior. 
It measures the frequency of each AS that originates a given prefix; if the frequency is 
lower than a specified threshold, then it could be a potential hijack attack.
(2) \emph{Time Heuristic.}  Most known attacks, including those discussed in \S~\ref{sec:hijacks_and_interception}, 
last a relatively short amount of time. The time heuristic measures the amount of time each 
path to a prefix is announced for; if the amount of time is extremely small (below a specified threshold), 
then there is the possibility of it being a routing attack.  

\emph{Detection Capability:} We tested our BGP monitoring framework based on BGP data during 
known prefix hijack attacks, that were discussed in \S~\ref{sec:hijacks_and_interception}. 
As a preliminary validation, the frequency and time heuristics were able to detect all of the known attacks; the threshold 
used for the frequency heuristic was .00001 and the threshold used for the time heuristic was .01. 

\textbf{Traceroute Monitoring Framework.}
The BGP monitoring framework provides measurements of actual AS-level paths from BGP collector nodes. 
However, the input data to the monitoring framework is limited to peers who chose to participate in 
frameworks such as Routeviews, and BGP data is only a noisy indicator of the routing control-plane. For robust 
detection of attacks, it is also necessary to monitor the data-plane, which we do via a Traceroute 
monitoring framework. 

Traceroute is a network diagnostic tool that infers the routers traversed by internet packets. To analyze 
both attacks and changes in AS-level paths to the Tor network, we have built a traceroute monitoring 
framework that runs traceroutes from  ~450 PlanetLab machines to all Tor entry and exit relays and stores the 
resulting traceroute data. The set of all Tor entry and exit relays is updated daily to accommodate new relays 
that have received the guard and exit flags. 
BGP hijack and interception attacks typically affect a variety of users from different vantage points. Thus, 
traceroute measurements from 450 geographically diverse PlanetLab have the ability to detect data-plane 
anomalies arising out of routing attacks. The PlanetLab machines are distributed across 140 ASes. Meanwhile, 
the Tor entry relays are distributed across 982 ASes and the exit relays are distributed across 882 ASes. 
We use Team-Cymru (\url{http://www.team-cymru.org/}) to compute the mapping between an IP address and its 
autonomous system. We will make the data collected by our Traceroute monitoring framework available to the 
research community.

\emph{Detection Capability:} As a preliminary validation, our Traceroute monitoring framework was able to 
detect the the BGP interception attack discussed in \S~\ref{sec:hijacks_and_interception}. From the traceroute 
data, we observed AS-level path changes from \emph{every} PlanetLab node to our Tor guard relay, indicating an anomaly.

\subsubsection{Preventing Routing Attacks in Tor}
In addition to monitoring the routing control-plane and data-plane with respect to the Tor network, the following 
approaches can help \emph{prevent} the threat of Raptor's routing attacks.

\noindent{\bf Advertising \texttt{/24} Tor prefixes:} Our experimental measurements indicate that over 90\% of Tor relays
 have a prefix length shorter than \texttt{/24}. This allows an AS-level adversary to launch a BGP hijack or 
interception attack against these Tor relays by advertising a more specific prefix for them (globally). We advocate that the 
Tor relay operators should be running Tor relays with a prefix length of \texttt{/24}. Autonomous systems typically filter route 
advertisements of prefix longer than \texttt{/24}, so AS-level adversaries will not be able to launch a more 
specific hijack or interception attack.

\noindent{\bf Favoring closer guard relays:} Even if a Tor relay advertises a \texttt{/24} prefix, an AS-level adversary can 
launch an equally specific prefix hijack  or interception attack (by advertising another \texttt{/24}). In this case, 
the impact of the attack is localized around the attacker's autonomous system, since the route is not globally propagated. 
We advocate that Tor clients select their guard relays by favoring Tor relays with a shorter AS-level path between them.
Tor clients could either obtain AS-level path information via the Tor network consensus download mechanism, or they can 
perform traceroutes themselves.
This further mitigates the risk to Tor clients due to an equally specific prefix attack. We note that by selecting guard 
relays that are closer to the client in the AS topology, the risk of asymmetric traffic analysis and BGP churn is also mitigated. 
\footnote{We note that if clients select closer guards, then knowledge of the guards reveals probabilistic information about the 
clients. We will investigate this trade-off in future work.}

\noindent{\bf Securing inter-domain routing:} The research community has proposed multiple protocols for securing 
inter-domain routing~\cite{oorschot:tissec07,hu:sigcomm04,chan:sigcomm06,gill:sigcomm11,boldyreva:ccs12}. Real-world 
deployment of these protocols would mitigate the BGP hijack and interception attacks on Tor. However, this 
approach requires buy-in from multiple stakeholders in the complex ecosystem of the Internet, and progress 
on this front has been slow. We hope that the concerns we raise about the compromise of user anonymity in Tor 
can help accelerate the momentum for improving BGP security. 

\section{Discussion and Ethical Considerations}
\label{sec:ethics}

\textbf{Colluding adversaries.} In this paper, we quantified the threat of Raptor 
attacks from the perspective of individual autonomous systems. In practice, autonomous 
systems can collude with each other to increase their capability of monitoring Tor traffic. 
For example, autonomous systems within the same legal jurisdiction may be forced 
to monitor Tor traffic and share it with a single entity that may launch Raptor attacks.  

\noindent\textbf{Applicability to other anonymity systems.} It is important to note that our attacks merely 
consider Tor as an example of a low-latency anonymity system. 
Raptor attacks are broadly applicable to other deployed anonymity systems such as I2P, Freenet and Tribler~\cite{i2p,freenet,tribler}.

\noindent \textbf{Ethical considerations.} We introduce and evaluate several novel attacks 
against the Tor network. The Tor network has a userbase of several million users~\cite{tor-users}, 
and these users are especially concerned about the privacy of their communications. 
Thus, it is of utmost importance that our real-world experiments on the Tor network do not 
compromise the privacy and safety of Tor users. In this paper, we take multiple precautions 
to safeguard the privacy of Tor users:

\begin{itemize}[leftmargin=*]
\setlength{\itemsep}{-1pt}
\item \emph{Attack our own traffic}.  All of our attacks only experiment with traffic that we created ourselves, 
i.e., we deanonymize our own traffic. In fact, we do not store or analyze traffic of any real Tor user.
\item \emph{Attack our own relay}. Similarly, to demonstrate the threat of prefix interception attacks on the live 
Tor network, we launch interception attacks against relays that we already control, i.e., we hijack/intercept our own prefix.
\item \emph{Firewall our Tor relay}. We also used network-level firewalls to ensure that real Tor users will never use relays 
that we control: traffic from real users is dropped by the firewall. Only authorized traffic that we create ourselves 
can bypass the firewall and use our Tor relay.  
\end{itemize}

\section{Related Work}

\noindent\textbf{AS-level adversaries:} It is well known that an adversary who can 
observe users' communications at both ends of the segment can deanonymize 
Tor clients~\cite{shmatikov:esorics06,zhu:pet04}. Feamster and Dingledine 
were the first to consider the attack from the perspective of an AS-level 
adversary~\cite{feamster:wpes04}. Later, Edman and Syverson explored the 
impact of Tor path selection strategies on the security of the 
network~\cite{edman:ccs09}. Recently, Johnson et al. analyzed the security 
of the Tor network against AS-level adversaries in terms of user understandable 
metrics for anonymity~\cite{johnson:ccs13}, and Akhoondi \emph{et al}.~\cite{akhoondi:sp12} 
considered path selection algorithms that minimize opportunities for AS-level end-to-end traffic analysis. 
Finally, Murdoch \emph{et al}.~\cite{murdoch:pet07} considered the analogous analysis with 
respect to Internet exchange level adversaries, which are also in a position 
to observe a significant fraction of Internet traffic.

We build upon these works and introduce Raptor attacks, that leverage routing 
asymmetry, routing churn, and routing attacks to compromise user anonymity more 
effectively than previously thought possible.  

The attack observations in Raptor were briefly discussed in a preliminary and short workshop paper~\cite{Vanbever:2014:AQU:2670518.2673869}. 
In this paper, we go further by measuring the importance of the attacks using real-world Internet control- and data-plane data. We also demonstrate 
the attacks feasibility by performing them on the live Tor network---\emph{with success}. Finally, we also describe efficient countermeasures to 
restore a good level of anonymity. %

\noindent\textbf{Traffic analysis of Tor:} An important thread of research aims to perform 
traffic analysis of Tor communications via side-channel 
information about Tor relays. Murdoch et al.~\cite{murdoch:sp05}, Evans et al.~\cite{evans:sec09}, and 
Jansen et al.~\cite{jansen:ndss14} have demonstrated attacks that use node congestion and protocol-level details 
as a side channel to uncover Tor relays involved in anonymous paths. Furthermore, Mittal et al.~\cite{mittal:ccs11} and 
Hopper et al.~\cite{hopper:ccs07,hopper:tissec10} 
proposed the use of network throughput and network latency as a side channel to fingerprint Tor relays involved in anonymous paths. 
We note that most of these attacks provide probabilistic information about Tor relays, and may not deanonymize the Tor clients. 
In contrast, Raptor attacks can completely deanonymize Tor clients. 

\noindent\textbf{BGP insecurity:} The networking research community has extensively studied attacks 
on inter-domain routing protocols including BGP hijack~\cite{zhang:conext07,zhang:ton10,zheng:sigcomm07,shi:imc12} 
and interception attacks~\cite{ballani:sigcomm07}. Similarly, there has been much work on proposing secure 
routing protocols that resist the above attacks~\cite{oorschot:tissec07,hu:sigcomm04,boldyreva:ccs12,chan:sigcomm06,gill:sigcomm11}. 
However, we are the first to study the implications of these attacks on privacy technologies such as the Tor network. 
Arnbak et al.~\cite{arnbak:hotpets14} discuss surveillance capabilities of autonomous systems from a legal perspective, 
but do not discuss anonymity systems.

\section{Conclusion}
Raptor attacks exploit the dynamics of Internet routing (such as routing asymmetry, 
routing churn, and routing attacks) to enable an AS-level adversary to effectively 
compromise user anonymity.

Our experimental results show that Raptor attacks present a serious threat to 
the security of anonymity systems. Our key results include (1) demonstration of 
asymmetric traffic correlation on the live Tor network, which achieves  95\% accuracy 
with no observed false positives, (2) quantifying the impact of 
routing asymmetry and routing churn on AS-level attacks -- an increase of 50\% 
to 100\% respectively compared to conventional attacks, (3) uncovering historical 
BGP hijacks involving Tor relays, and (4) successful demonstration of a traffic analysis 
attack via BGP interception  on the live Tor network. We also outlined a taxonomy of 
countermeasures against our attacks. 

Our work highlights the dangers of abstracting network routing from the analysis 
of anonymity systems such as Tor, and motivates the design of next generation anonymity 
systems that resist Raptor.

\section{Acknowledgments}

Thanks to Ethan Katz-Bassett for support on setting up Transit Portal provided by the PEERING project. Thanks to ATLAS project for donating credits for our experimental setup. Thanks to Matthew Wright, Nick Feamster, Nikita Borisov and Roger Dingledine for helpful discussions. This work was supported by the NSF under the grant CNS-1423139.

\bibliographystyle{acm}
\bibliography{local,as}

\appendix
\section{Appendix}

\subsection{Monitoring Challenges for Detecting 
Routing Attacks}
\label{appendix:monitoring-challenges}
Detecting a malicious (or accidental) event, such as a prefix hijack or interception 
could be very helpful to a Tor user, but it also introduces a new possible attack. If users can select Tor relays that 
they know are not compromised, then a clever attacker could hijack all Tor relays except for a few; then the user would 
be forced to use a small set of relays, which would be known to the attacker.  While this vulnerability is introduced, 
our detection heuristics still bring accountability to the Tor network.  Additionally, the Tor network can still use 
these detection mechanisms because of the short life span of hijack attacks; authorities can suspend the use of 
certain relays for a short amount of time, which should not significantly affect the Tor network.  

A difficult problem in hijack detection stems from the fundamental issues of BGP; many times events are flagged as 
suspicious, but in reality are a result of multiple origin ASes, traffic engineering, or load-balancing.  
Prior research in the area of detecting prefix hijack and interception attacks had the primary goal of providing 
no false negatives and very few false positives; our goal is slightly different.  We also have the goal of providing 
no false negatives, but false positives are less important; our goal is to increase anonymity over accuracy. 
This coupled with the ability to suspend relays for short periods of time (even for a false positive), 
should not cause harm, and should increase anonymity.

\subsection{Rejected countermeasure: Mitigating Correlation Attacks}
\label{appendix:obfuscation}

\noindent{\bf Obfuscating packet timings and sizes:} %
While the use of high latency mix networks~\cite{mixmaster,danezis:sp03} and constant rate 
cover traffic~\cite{restricted-routes} can mitigate timing analysis even against an adversary 
that observes all communications, these defenses are considered too costly to be deployed in 
the Tor network.

\noindent{\bf Mitigating asymmetric attacks:} Recall that our asymmetric correlation attack leverages information
in the TCP header, namely the sequence number field that indicates the number of acknowledged bytes. 
One potential 
countermeasure would be to encrypt the TCP header, by leveraging IP-layer encryption techniques such 
as IP-Sec. However, this approach introduces several challenges. First, it would require a substantial 
engineering effort to migrate Tor towards IPSEC. Second, since IPSEC is not widely used, this would make 
Tor traffic easy to distinguish from other encrypted traffic, thwarting its use for applications such 
as censorship resistance. Finally, encrypting the TCP header may not complete solve the attack. For 
example, an adversary could attempt to correlate TCP data packets with simply the number of TCP Ack packets, 
disregarding the sequence number field. 

\end{document}